\documentclass[]{fairmeta}

\let\txtaccentL\L
\let\txtaccentr\r

\usepackage[disable,textsize=tiny]{todonotes}
\usepackage{xcolor}
\usepackage{colortbl}
\usepackage{xspace}
\usepackage{siunitx}
\usepackage{graphicx}
\usepackage{tikz}
\usetikzlibrary{arrows.meta,positioning,shapes.geometric,calc,fit,backgrounds,decorations.pathreplacing}
\definecolor{accentblue}{RGB}{42,111,219}
\definecolor{darktext}{RGB}{26,26,46}
\usepackage{tabularx}
\usepackage{makecell}
\usepackage{algorithm,algorithmic}
\usepackage{paralist,amsmath, amssymb,bm}
\usepackage{multirow}
\usepackage{ntheorem}

\usepackage{enumitem}
\usepackage{changepage}
\usepackage{textcase}
\usepackage{booktabs}
\usepackage{placeins}

\newcommand{\eqnref}[1]{Equation~\eqref{#1}}

\newcommand{\scen}[1]{\textsc{#1}}
\newcommand{\armname}[1]{\textsc{#1}}

\usepackage{fancybox}
\usepackage{mathtools}
\usepackage{pifont}

\newlength\savewidth

\newcolumntype{x}[1]{>{\centering\arraybackslash}p{#1pt}}

\let\cedillaaccent\c

\def \L {\mathcal{L}}

\def \R {\mathbb{R}}

\def \x {\mathbf{x}}
\def \mX {\mathbf{X}}

\def \p {\mathbf{p}}

\def \c {\ifmmode\mathbf{c}\else\expandafter\cedillaaccent\fi}

\def \z {\mathbf{z}}

\def \g {\mathbf{g}}

\def \r {\mathbf{r}}

\def \I {\mathcal{I}}

\def \f {\mathbf{f}}
\def \h {\mathbf{h}}

\graphicspath{{figures/}}

\newcommand{\methodname}{CAP-Mix\xspace}

\newcommand{\clustids}{prototypes\xspace}

\newcommand{\pending}[1]{\textcolor{red}{#1}}

\newcommand{\QoneVsFlatBeauty}{$+2.08\%$}
\newcommand{\QoneVsFlatIndus}{$+4.73\%$}
\newcommand{\QoneVsFlatSteam}{$+0.72\%$}

\newcommand{\QtwobVsFlatBeauty}{$+1.49\%$}
\newcommand{\QtwobVsFlatIndus}{$-0.52\%$}
\newcommand{\QtwobVsFlatSteam}{$+2.09\%$}

\newcommand{\HardVsFlatPcaBeauty}{$-3.89\%$}
\newcommand{\HardVsFlatPcaIndus}{$-7.61\%$}

\newcommand{\QtwoaRangeLo}{$4.45\%$}
\newcommand{\QtwoaRangeHi}{$14.90\%$}
\newcommand{\FrozenRangeLo}{$14.15\%$}
\newcommand{\FrozenRangeHi}{$20.21\%$}

\newcommand{\NoPriorSidSteam}{$+6.91\%$}

\newcommand{\AvgHRfive}{$+6.9\%$}
\newcommand{\AvgNDCGfive}{$+13.6\%$}

\AtBeginDocument{%
  \def\L{\ifmmode\mathcal{L}\else\txtaccentL\fi}%
  \def\r{\ifmmode\mathbf{r}\else\txtaccentr\fi}%
}

\title{Item Representations Matter: Bridging Hard Semantic IDs and Soft Prototype Mixtures for Generative Retrieval}

\author[1,2,*]{Vicente Balmaseda}
\author[1]{Dongqi Fu}
\author[1]{Jiarui Feng}
\author[1]{Ren Chen}
\author[1]{Hanqing Zeng}
\author[1]{Tianxin Wei}
\author[1]{Yinglong Xia}
\author[2]{Tianbao Yang}
\author[1]{Hong Li}

\affiliation[1]{Meta}
\affiliation[2]{Texas A\&M University}

\contribution[*]{Work done at Meta}

\abstract{%
Modern recommender systems rely on generative retrieval techniques to predict the next item either as discrete codes or as a continuous vector.
On the one hand, Semantic IDs (SIDs) give discrete models shared parameters and pretrained semantic structure, but discretizing items into hard codes discards some information from the original content embeddings.
On the other hand, continuous models avoid quantization, yet typically use flat item lookup tables with no explicit parameter tying across items.
We ask whether a continuous recommender can introduce shared structure without making it a discrete bottleneck.
We propose \textbf{\methodname}, a flow recommender that represents each item as \textit{a soft mixture of shared prototypes} plus \textit{an item-specific vector}.
The assignments, prototypes, and item vectors are trained jointly, and the soft representation is retained at inference.
Pretrained information is optional: SID responsibilities can serve as a prior over assignments, while a full content embedding can initialize the item-specific vector without changing the initial flat table.
In matched comparisons on three datasets, \methodname\ improves over a flat table on the two sparse Amazon datasets and yields a small gain on Steam when trained without pretrained information. Given only an offline SID, it outperforms a hard table built from the same tokenizer by \QtwoaRangeLo--\QtwoaRangeHi\ in HR@5~$\uparrow$.
Given the full content embedding, it matches or improves a strong flat table while hard assignments regress on the sparse Amazon datasets.
Overall, averaged across six benchmarks, \methodname\ improves HR@5~$\uparrow$ and NDCG@5~$\uparrow$ by \AvgHRfive\ and \AvgNDCGfive\ relative to the best evaluated baseline for each dataset and metric, while delivering $1.4$--$161\times$ the inference throughput~$\uparrow$ of the discrete baselines in our experiments.%
}

\correspondence{Vicente~Balmaseda (\email{vibalcam@tamu.edu}), Dongqi~Fu (\email{dongqifu@meta.com})}

\begin{document}

\maketitle

\section{Introduction}
\label{sec:intro}

Generative retrieval techniques enable modern recommender systems to perform large-scale next-item retrieval by learning the next-item distribution and generating recommendations from it, using either discrete tokens~\citep{rajput2023tiger} or continuous item embeddings~\citep{li2025preferencetrajectorymodelingflow}.
Either way, a recommender needs an item representation table (or lookup table) as input, i.e., the map from a catalog to the representations it generates or scores against, and how that map is built is a lever in its own right, open to external information.
Traditionally, it is \emph{flat}: one independently parameterized
vector per item, with no explicit parameter tying between item rows. Recently, discrete methods have replaced that default, whereas continuous methods have retained it.

\emph{Discrete} models tokenize each item embedding into codes and emit those codes autoregressively~\citep{deng2025onerecunifyingretrieverank,rajput2023tiger} or by masked diffusion~\citep{shah2025madrec}. They owe much of their strength to those codes being Semantic IDs (SIDs)~\citep{singh2024better}. A residual quantizer~\citep{Oord2017,LeeKKCH22,rqgmm2026} turns each item's pretrained content embedding into a tuple of codes, so the item vocabulary becomes a clustering~\citep{grid}.
Two things follow: items sharing a code share the parameters attached to it, so their gradients are aggregated through the shared codeword, introducing explicit cross-item coupling and thereby pooling training signal across items; and because the codes are quantized from a content embedding, that grouping imports semantic structure.
However, tokenization has two costs: committing each item to a
single codeword at each level generally introduces reconstruction error, and distinct items may also collide on the same code tuple. A recent line of work makes the tokenizer itself adaptive by regularizing it~\citep{10.1145/3627673.3679569}, coupling it to the recommender~\citep{10.1145/3726302.3729989}, or relaxing the assignment so gradients pass through it~\citep{li2026unigrecunifiedgenerativerecommendation,10.1145/3805712.3809641}, but recommendation is still organized around emitting discrete codes, and both costs remain: they stem from
tokenization itself, not from when the tokenizer is fit.

\emph{Continuous} models instead represent each item as a vector in $\mathbb{R}^d$ and retrieve those nearest to the one they
predict~\citep{10.1145/3539618.3591663,10.24963/ijcai.2025/346,li2025preferencetrajectorymodelingflow}. They do not tokenize and are free to seed the content embedding straight into their item table, so semantic structure reaches them without any codes. What they do not obtain automatically is explicit cross-item parameter sharing: every item row remains independently parameterized.
This raises our central question: can a continuous generative recommender obtain adaptive cross-item parameter sharing without converting items into discrete tokens?

A simple way to use an SID in a continuous recommender is to replace its flat lookup table with a hard SID
table whose assignments are frozen. If the tokenizer assigns item $i$ the code $c_i^{(s)}$ at level $s$, then its embedding is
$\x_i=\sum_s C^{(s)}[c_i^{(s)}] + \r_i$, where $C^{(s)}$ contains the codeword vectors and the item-specific vector $\r_i$ can be initialized from the original content embedding. This preserves continuous retrieval while transferring the SID's parameter sharing: items with the same code update a shared vector.
However, our diagnostic comparison (Appendix~\ref{app:diagnostics}, Table~\ref{tab:motivation}) shows that, without content embeddings, semantic assignments outperform matched random assignments, indicating that the imported structure is useful. When the full content embedding initializes $\r_i$, by contrast, frozen hard semantic assignments hurt, whereas random assignments remain near the flat baseline.
These observations suggest that sharing can help, but that a frozen semantic partition can become restrictive when its underlying continuous representation is already available. This
motivates shared structure that adapts jointly with the recommender while retaining item-specific degrees of
freedom.

Following these observations, we propose \methodname, a flow recommender whose item table is a co-adaptive
mixture of shared \clustids\ plus an item-specific vector.
Rather than selecting one codeword per level, an item \emph{softly} mixes over
all of them, so the grouping is differentiable and the ranking objective moves it continuously instead of by
discrete reassignment, while the item-specific vector keeps it from being reduced to the group it lands in.
We call this a \emph{prototype-mixture} item table:
\begin{equation}
\x_i \;=\;  \frac{\gamma}{S}\sum_{s=1}^{S} A^{(s)}_i C^{(s)} + \r_i,
\qquad \sum_{k=1}^{K} A^{(s)}_i[k] \;=\; 1,
\label{eq:intro}
\end{equation}
with $C^{(s)}$ the $K$ shared \clustids of level $s$, $A^{(s)}_i$ the weights item $i$ places on them,
$\r_i$ the item's own vector, and $\gamma>0$ a scale factor, all learned with the flow in a single stage.

Empirically, we evaluate \methodname\ across six public benchmarks (Table~\ref{tab:main-summary}), where \methodname\ improves average HR@5~$\uparrow$ and NDCG@5~$\uparrow$ by \AvgHRfive\ and \AvgNDCGfive, respectively, relative to the strongest evaluated baseline (FlowRec, TIGER, or MaskGR) for each dataset and metric, while delivering $1.4$--$161\times$ the inference throughput~$\uparrow$ of the evaluated discrete baselines (Table~\ref{tab:cost-summary}).

\section{Preliminaries}
\label{sec:prelim}

\subsection{Flow-based generative retrieval for recommender systems}
\label{sec:fm}

Flow matching~\citep{lipman_flow_2023} learns to transform samples from a simple source distribution
$p_{\mathrm{source}}$ into samples from a target distribution $p_{\mathrm{data}}$. It learns a velocity field
$v_\theta(\x,t)$ that tells each sample how to move at time $t$.
Let $\I$ be the item catalog, $H=(i_1,\ldots,i_m)$ a user's interaction history, and $i_{m+1}$ the next item.
A flow-based generative recommender can be decomposed into three components. First, the item encoder
$E_\theta:\I\rightarrow\R^D$ maps each item $i$ to an embedding $\x_i$. Applied elementwise to $H$, it yields
$(\x_{i_1},\ldots,\x_{i_m})$. Second, the flow starts from a source $\x_0$ and follows the learned
velocity field $v_\theta(\x,t)$ toward a predicted next-item embedding by integrating from $t=0$ to $t=1$.
Writing $v$ for the exact velocity field, explicit Euler approximates this transport using $N$ evaluations of the
learned field:
\begin{equation}
    \hat\x_1
    =\x_0+\int_0^1 v(\x_t,t)\,\mathrm{d}t
    \approx \x_0+h\sum_{n=0}^{N-1}v_\theta(\hat\x_{t_n},t_n),
    \qquad t_n=nh,\quad h=1/N.
\label{eq:ode}
\end{equation}
Here, $\hat\x_{t_n}$ denotes the numerical state at time $t_n$, with $\hat\x_{t_0}=\x_0$. Depending on the
architecture, the encoded history may determine the source, condition the velocity field, or both.
Third, the decoder $D_\theta(\hat\x_1,H)\in\Delta^{|\I|-1}$ converts this endpoint into a
distribution over the catalog by scoring it against the item table
$\mX=[\x_i^\top]_{i\in\I}\in\R^{|\I|\times D}$. Figure~\ref{fig:arch}(a)
illustrates this complete encoder--flow--decoder path.
During training, the target endpoint is the next-item embedding $\x_1=\x_{i_{m+1}}$. For a source $\x_0$ and
target $\x_1$, conditional flow matching commonly uses the straight path and regresses the conditional velocity
\begin{equation}
\x_t=(1-t)\x_0+t\x_1,\qquad
v^\star=\x_1-\x_0,\qquad
\L_{\mathrm{CFM}}
=
\mathbb{E}_{t\sim\mathcal{U}[0,1]}
\left\|v_\theta(\x_t,t)-v^\star\right\|_2^2 .
\label{eq:cfm}
\end{equation}
The same item table therefore embeds the history, defines the flow's training targets, and supports retrieval.
Existing flow-based recommenders~\citep{10.24963/ijcai.2025/346,li2025preferencetrajectorymodelingflow} focus
primarily on the transport, whereas we study how the item-table parameterization can enable sharing across items.

\subsection{Semantic IDs}
\label{sec:sid}

A Semantic ID (SID) represents an item by an ordered tuple of codes, typically obtained by applying residual
quantization to a pretrained content embedding~\citep{rajput2023tiger,grid}. Given $L$ codebooks
$Q^{(\ell)}\in\R^{K\times D}$, residual quantization decomposes the content embedding $\mathbf e_i$ into selected
codewords and a final residual $\boldsymbol\varepsilon_i$:
\begin{equation}
\operatorname{SID}(i)=\bigl(c_i^{(1)},\ldots,c_i^{(L)}\bigr),
\qquad
\mathbf e_i=\sum_{\ell=1}^{L}Q^{(\ell)}[c_i^{(\ell)}]+\boldsymbol\varepsilon_i,
\qquad c_i^{(\ell)}\in\{1,\ldots,K\}.
\label{eq:sid}
\end{equation}
The SID retains the selected indices but not $\boldsymbol\varepsilon_i$. Methods differ in how they obtain these
assignments: RQ-KMeans selects the nearest centroid~\citep{grid}, RQ-VAE learns its quantization space through
reconstruction~\citep{LeeKKCH22}, and RQ-GMM assigns items using posterior mixture
probabilities~\citep{rqgmm2026}. Regardless of the fitting procedure, the SID used at inference is \emph{hard}: each
level selects one codeword, whose index becomes that level's code. Items that share a code reuse its parameters,
thereby pooling recommendation gradients across their interactions. Deriving the codes from content embeddings also
transfers pretrained semantic structure.

Recent methods remove the need for a separately pretrained, frozen tokenizer by adapting it jointly with the
recommender~\citep{li2026unigrecunifiedgenerativerecommendation,10.1145/3805712.3809641,wang2026pitdynamicpersonalizeditem},
but their deployed IDs remain hard and discard the residual $\boldsymbol\varepsilon_i$. SIDs therefore share
parameters through discrete code assignments at the cost of quantizing the item representation. Flat-table
continuous recommenders avoid quantization but learn each item embedding independently. Our goal is to combine
adaptive sharing with trainable continuous representations.

\section{Method: \methodname}
\label{sec:method}

Following Section~\ref{sec:fm}, \methodname\ uses a continuous flow to predict a next-item embedding and a catalog
decoder to model $p(i\mid H)$. Its main contribution is the \textbf{item encoder} $E_\theta$, which replaces a flat
item table with a \emph{prototype-mixture} parameterization. This representation learns soft item groupings for
cross-item sharing while retaining an item-specific vector, and can incorporate pretrained SIDs or content embeddings
when available. Figure~\ref{fig:arch} provides an overview.

\definecolor{cFeat}{RGB}{214,218,226}   %
\definecolor{cBkA}{RGB}{124,196,170}    %
\definecolor{cBkB}{RGB}{152,134,206}    %
\definecolor{cResid}{RGB}{247,192,124}  %
\definecolor{cPanel}{RGB}{247,249,253}  %

\newcommand{\gstrip}[2]{%
  \tikz[baseline=-2.9pt, sharp corners]{\foreach \i in {1,...,#1}
    {\draw[fill=#2, draw=black!40, line width=0.2pt]
      ({(\i-1)*4.8pt},-2.9pt) rectangle ++(4.6pt,5.8pt);}}%
}
\newcommand{\gmix}[1]{%
  \tikz[baseline=-2.9pt, sharp corners]{\foreach \i in {1,...,#1}
    {\fill[cBkA]   ({(\i-1)*4.8pt}, 0.97pt) rectangle ++(4.6pt, 1.93pt);
     \fill[cBkB]   ({(\i-1)*4.8pt},-0.97pt) rectangle ++(4.6pt, 1.94pt);
     \fill[cResid] ({(\i-1)*4.8pt},-2.90pt) rectangle ++(4.6pt, 1.93pt);
     \draw[draw=black!40, line width=0.2pt]
       ({(\i-1)*4.8pt},-2.9pt) rectangle ++(4.6pt,5.8pt);}}%
}
\newcommand{\gbars}[2]{%
  \tikz[baseline=-1pt, sharp corners]{%
    \draw[black!30, line width=0.25pt] (-1pt,0) -- (20.5pt,0);
    \foreach \h [count=\i] in #2
      {\draw[fill=#1, draw=#1!65!black, line width=0.2pt]
        ({(\i-1)*4.1pt},0) rectangle ++(3.0pt,{\h*12pt+1.0pt});}%
  }%
}
\newcommand{\gbook}[2]{%
  \tikz[baseline=7pt, sharp corners]{%
    \foreach \h [count=\i] in #2
      {\draw[fill=white, draw=black!35, line width=0.2pt]
         (0,{-(\i-1)*3.9pt}) rectangle ++(24pt,-3.0pt);
       \fill[#1, opacity={0.14+0.86*\h}]
         (0.25pt,{-(\i-1)*3.9pt-0.25pt}) rectangle ++(23.5pt,-2.5pt);}%
  }%
}

\begin{figure}[t]
\centering
\begin{minipage}[c]{0.27\linewidth}\centering
\begin{tikzpicture}[
  font=\scriptsize,
  cbox/.style  ={draw=black!45, rounded corners=1.6pt, inner sep=2.6pt, align=center,
                 line width=0.45pt, fill=white, minimum width=15mm},
  plain/.style ={inner sep=1.5pt, align=center},
  ar/.style    ={-{Latex[length=3.6pt,width=3.0pt]}, black!65, line width=0.45pt},
  dar/.style   ={-{Latex[length=3.2pt,width=2.6pt]}, black!42, line width=0.4pt,
                 dash pattern=on 1.6pt off 1.4pt},
  note/.style  ={inner sep=1pt, text=black!58, font=\tiny, align=center, fill=white},
  ttl/.style   ={font=\scriptsize\bfseries, text=accentblue, inner sep=2pt, fill=white},
]
\node[plain] (hin) at (0, 1.62) {history $H$};
\node[cbox, fill=cFeat!45]  (enc) at (0, 0.92) {encoder $E_\theta$};
\node[cbox, fill=cBkA!30]   (flw) at (0, 0.02) {flow\\[-1pt]\tiny $v_\theta(\x_t,t)$};
\node[cbox, fill=cBkB!28]   (dec) at (0,-0.94) {decoder $D_\theta$};
\node[plain] (out) at (0,-1.68) {top-$k$ items};
\draw[ar] (hin) -- (enc);
\draw[ar] (enc) -- node[right, note] {$\x_0$} (flw);
\draw[ar] (flw) -- node[right, note] {$\hat\x_1$} (dec);
\draw[ar] (dec) -- (out);

\draw[ar] (flw.-20) to[out=-12,in=12,looseness=1.55]
  node[right, note, xshift=-1pt] (repeatlabel) {$N\!\times$} (flw.20);

\node[cbox, fill=cResid!35, minimum width=0pt, inner sep=2.2pt]
  (tab) at (-1.55, 0.03) {\rotatebox{90}{item table $\{\x_i\}$}};
\draw[dar] (tab.east) to[out=0,in=180] (enc.west);
\draw[dar] (tab.east) to[out=0,in=180] (dec.west);
\begin{scope}[on background layer]
  \node[draw=accentblue!55, fill=cPanel, rounded corners=3pt,
        dash pattern=on 2.2pt off 1.8pt, line width=0.5pt,
        fit=(hin)(enc)(flw)(dec)(out)(tab)(repeatlabel), inner sep=4pt] (Ppipe) {};
\end{scope}
\node[ttl, anchor=west] at ($(Ppipe.north west)+(5pt,0)$) {(a) Pipeline};
\end{tikzpicture}
\end{minipage}\hfill
\begin{minipage}[c]{0.71\linewidth}\centering
\begin{tikzpicture}[
  scale=0.89, every node/.style={transform shape},
  font=\scriptsize,
  box/.style   ={draw=black!45, rounded corners=1.6pt, inner sep=3pt, align=center,
                 line width=0.45pt, fill=white},
  plain/.style ={inner sep=1.5pt, align=center},
  op/.style    ={draw=black!55, fill=white, circle, inner sep=0pt, minimum size=3.2mm,
                 line width=0.45pt},
  ar/.style    ={-{Latex[length=3.6pt,width=3.0pt]}, black!65, line width=0.45pt},
  note/.style  ={inner sep=1.2pt, text=black!58, font=\tiny, align=center, fill=white},
  ttl/.style   ={font=\scriptsize\bfseries, text=accentblue, inner sep=2pt, fill=white},
]

\node[plain] (it) at (-0.10, 0) {item $i$};
\node[box] (f) at (1.25, 0) {\gstrip{4}{cFeat}\\[2pt]$\f_i$};
\draw[ar] (it) -- (f);

\node[box] (A1) at (3.05, 1.05) {\gbars{cBkA}{{0.09,0.24,1.00,0.17,0.06}}\\[2pt]$A^{(1)}_i$};
\node[op]  (m1) at (4.25, 1.05) {$\scriptscriptstyle\times$};
\node[box] (C1) at (5.55, 1.05) {\gbook{cBkA}{{0.09,0.24,1.00,0.17,0.06}}\\[2pt]$C^{(1)}$};
\node[box] (AS) at (3.05,-1.05) {\gbars{cBkB}{{0.07,0.14,0.21,0.95,0.11}}\\[2pt]$A^{(S)}_i$};
\node[op]  (mS) at (4.25,-1.05) {$\scriptscriptstyle\times$};
\node[box] (CS) at (5.55,-1.05) {\gbook{cBkB}{{0.07,0.14,0.21,0.95,0.11}}\\[2pt]$C^{(S)}$};
\draw[ar] (A1) -- (m1);  \draw[ar] (m1) -- (C1);
\draw[ar] (AS) -- (mS);  \draw[ar] (mS) -- (CS);
\node[plain] at (3.05,0) {$\vdots$};
\node[plain] at (5.55,0) {$\vdots$};

\draw[ar] (f.east) -- (A1.west);
\draw[ar] (f.east) -- (AS.west);
\node[note] at (2.20,0.02) {softmax\\[-0.5pt]$\phi_s(\cdot)$};

\node[box] (sm) at (7.00,0) {$\dfrac{\gamma}{S}\displaystyle\sum_{s=1}^{S}$};
\draw[ar] (C1.east) -- (sm.160);
\draw[ar] (CS.east) -- (sm.200);

\node[op] (pl) at (8.05,0) {$\scriptscriptstyle+$};
\node[box] (ri) at (8.05,-1.46)
  {\gstrip{8}{cResid}\\[2pt]$\r_i$\,\textcolor{black!58}{\tiny identity}};
\draw[ar] (sm) -- (pl);
\draw[ar] (ri) -- (pl);

\node[box] (xi) at (9.35,0) {\gmix{8}\\[2pt]$\x_i$};
\draw[ar] (pl) -- (xi);

\begin{scope}[on background layer]
  \node[draw=accentblue!55, fill=cPanel, rounded corners=3pt,
        dash pattern=on 2.2pt off 1.8pt, line width=0.5pt,
        fit=(it)(f)(A1)(AS)(C1)(CS)(sm)(ri)(xi), inner sep=5pt] (Penc) {};
\end{scope}
\node[ttl, anchor=west] at ($(Penc.north west)+(6pt,0)$) {(b) Item encoder $E_\theta$};

\end{tikzpicture}
\end{minipage}
\caption{
  \textbf{(a)} A general flow-based retrieval pipeline has three components, and the item table is
  shared across them: the encoder builds it, the flow is trained to land on one of its rows, and the decoder retrieves against it.
  \textbf{(b)} \methodname\ mainly changes the encoder of the pipeline: an attention block over the $S$ level tokens maps each item's short feature $\f_i\in\R^{D_f}$ to a \emph{soft} assignment $A^{(s)}_i$ over the $K$ \clustids of level $s$, which then blends the rows of that level's codebook $C^{(s)}$.
  Adding a trainable identity vector $\r_i$ yields the item embedding $\x_i\in\R^{D}$.
  Nothing is quantized or frozen, so the clustering is shaped by the training objective and the assignments are kept soft during training and inference.
}
\label{fig:arch}
\end{figure}
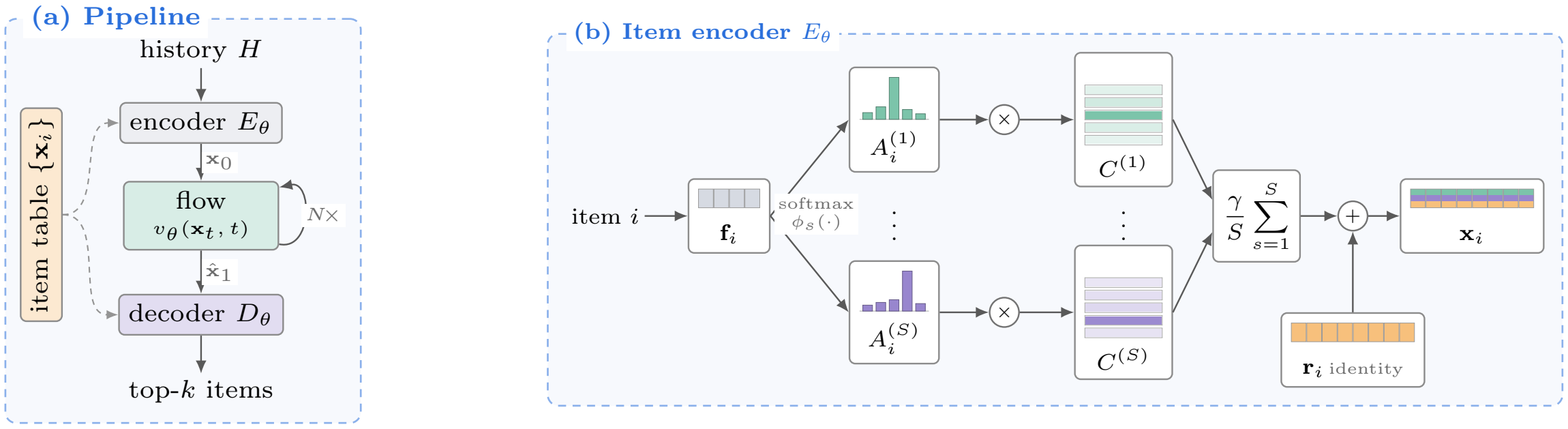

\subsection{Prototype-mixture item representations}
\label{sec:softcb}

\methodname\ maintains $S$ shared codebooks $C^{(s)}\in\R^{K\times D}$, whose $K$ rows are $D$-dimensional
prototype vectors. For item $i$, $A_i^{(s)}\in\Delta^{K-1}$ is a soft assignment over the prototypes in codebook $s$,
and $\r_i\in\R^D$ is an item-specific vector. The item embedding is then
\begin{equation}
\x_i
=
\underbrace{\vphantom{\frac{\gamma}{S}\sum_{s=1}^{S}A_i^{(s)}C^{(s)}}\r_i}_{\text{identity}}
+
\underbrace{\frac{\gamma}{S}\sum_{s=1}^{S}A_i^{(s)}C^{(s)}}_{\text{cross-item sharing}},
\qquad
A_i^{(s)}[k]\geq 0,\quad
\sum_{k=1}^{K}A_i^{(s)}[k]=1,
\label{eq:compose}
\end{equation}
where $A_i^{(s)}C^{(s)}\in\R^D$ is a convex mixture of the prototypes, $\gamma>0$ controls the contribution of the
shared component, and $1/S$ keeps its scale stable as $S$ changes.

\textbf{Cross-item sharing and item-specific memorization.}
Consider a differentiable objective $\mathcal L(\mX)$ whose dependence on the codebooks and item-specific
vectors is only through the composed item table $\mX$, and let $\g_i=\nabla_{\x_i}\mathcal L$. Differentiating
\eqnref{eq:compose} gives
\begin{equation}
\nabla_{C^{(s)}[k]}\mathcal L
=
\frac{\gamma}{S}\sum_{i\in\I}A_i^{(s)}[k]\g_i,
\qquad
\nabla_{\r_i}\mathcal L=\g_i.
\label{eq:shared-gradient}
\end{equation}
The prototype channel aggregates recommendation gradients across items by their assignment weights, providing
parameter sharing that can improve generalization. Meanwhile, $\r_i$ receives the unrestricted gradient $\g_i$,
preserving item-specific memorization and uniqueness even when assignments coincide.
Appendix~\ref{app:gradient} derives the resulting coupling under fixed assignments.

\textbf{Relation to flat and hard item tables.}
\eqnref{eq:compose} contains flat and hard item tables as special cases: $\gamma=0$ yields the flat
$\x_i=\r_i$, while fixed one-hot $A_i^{(s)}$ selects one prototype per codebook as in a hard SID lookup. Additionally setting
$\r_i=\mathbf 0$ gives the fully quantized case and discards item information not captured by the selected
codewords. \methodname\ instead keeps assignments soft and retains $\r_i$ for item-specific flexibility.

\textbf{Learning soft assignments.}
Each item has a learned low-dimensional feature $\f_i\in\R^{D_f}$. We derive one token per codebook from $\f_i$ and process the $S$
tokens jointly with a lightweight attention block:
\begin{equation}
\begin{aligned}
\z_i^{(s)}
&=W_{\mathrm{in}}^{(s)}\f_i+\p^{(s)},
&
(\h_i^{(1)},\ldots,\h_i^{(S)})
&=\operatorname{Attn}(\z_i^{(1)},\ldots,\z_i^{(S)}),\\
\phi_s(\f_i)
&=W_{\mathrm{out}}^{(s)}\h_i^{(s)},
&
A_i^{(s)}
&=\operatorname{softmax}\!\left(\phi_s(\f_i)\right).
\end{aligned}
\label{eq:assign}
\end{equation}
Here, $W_{\mathrm{in}}^{(s)}\f_i+\p^{(s)}$ is a codebook-specific affine projection, $\h_i^{(s)}$ is its
contextualized output, and $W_{\mathrm{out}}^{(s)}$ maps this output to $K$ assignment logits. This
parameterization has three properties. First, because $D_f<D$, $\f_i$ serves as an assignment bottleneck: it stores
compact information for selecting shared prototypes, while detailed item-specific information remains in $\r_i$.
Second, attention lets the assignments interact: it is bidirectional across codebooks by default and, when an
offline residual SID provides a level ordering, causally masked to respect that ordering. Third, no $\arg\max$ or straight-through estimator is used: the assignments
remain soft and differentiable during training and inference, and all representation parameters are learned jointly
with the recommender.

\subsection{Using optional pretrained information}
\label{sec:pretrained}

Pretrained information is optional: \methodname\ can learn entirely from recommendation interactions or incorporate
an offline SID or a full content embedding when either is available.

\textbf{From scratch.}
\methodname\ requires neither a tokenizer nor content embeddings. All representation parameters are randomly
initialized and learned solely from recommendation interactions. The resulting item groupings are therefore induced
by the recommendation objective rather than imported from pretrained content information.

\textbf{Offline Semantic ID.}
An offline residual tokenizer supplies both a distribution $\rho_i^{(t)}\in\Delta^{K-1}$ over codes at each depth
$t$ and the level ordering induced by its residual construction. We follow this ordering with $S$ parallel causal
chains of depth $T$, sharing the tokenizer prior at each depth across chains:
\begin{equation}
A_i^{(s,t)}
=\operatorname{softmax}\!\left(\phi_{s,t}(\f_i)+\alpha_t\log\!\left(\rho_i^{(t)}+\epsilon_\rho\right)\right),
\qquad
\x_i=\r_i+\frac{\gamma}{S}\sum_{s,t}A_i^{(s,t)}C^{(s,t)}.
\label{eq:sidprior}
\end{equation}
Here, the sum ranges over $s\in\{1,\ldots,S\}$ and $t\in\{1,\ldots,T\}$, and $\alpha_t\geq0$ is a learned trust
parameter. Because $\epsilon_\rho>0$ gives the prior full support, for any finite $\alpha_t$ the learned logits can
realize any distribution in the interior of the simplex. The assignments therefore remain soft and differentiable.
\eqnref{eq:compose} is recovered when $T=1$.
Appendix~\ref{app:priors} gives the pooling derivation and causal attention structure.

\textbf{Full content embedding.}
Let $\bar{\mathbf e}_i\in\R^D$ be the PCA-projected and coordinatewise-standardized content embedding. We initialize the item-specific
vector from this embedding and set every prototype to zero:
\begin{equation}
\r_i\leftarrow\bar{\mathbf e}_i,
\qquad
C^{(s)}\leftarrow\mathbf 0.
\label{eq:content-init}
\end{equation}
Both $\r_i$ and $C^{(s)}$ remain trainable after initialization. By \eqnref{eq:compose},
$\x_i=\bar{\mathbf e}_i$ initially, so \methodname\ and the flat content-initialized baseline begin from the same
unquantized item embeddings. The prototype channel then learns cross-item sharing from this common starting point.

\subsection{Flow backbone and decoder}
\label{sec:flow-decoder}

We build on the FlowRec backbone~\citep{li2025preferencetrajectorymodelingflow}, retaining its behavior
encoder, velocity field, and base flow objective while replacing its item encoder and decoder.
Appendix~\ref{app:impl} provides the full specification. A behavior encoder $B_\theta$ maps the $m$ most recent item embeddings to a history-dependent source,
while the next-item embedding supplies the target:
\begin{equation}
\x_0
=
B_\theta\!\left(E_\theta(i_1),\ldots,E_\theta(i_m)\right),
\qquad
\x_1=E_\theta(i_{m+1}).
\label{eq:prior}
\end{equation}
We train $v_\theta$ on the straight interpolant using the conditional flow-matching loss in
\eqnref{eq:cfm}. Following FlowRec, we also form the one-step endpoint estimate
\begin{equation}
\widetilde{\x}_1=\x_t+(1-t)v_\theta(\x_t,t).
\label{eq:onestep}
\end{equation}

\textbf{Decoder.}
For a query $\mathbf u\in\R^D$, the decoder combines dot-product scores against the shared item table with a
history-dependent co-visitation prior through log-linear pooling:
\begin{equation}
D_\theta(\mathbf u,H)[i]
=
\operatorname{softmax}_{i\in\I}\!\left(
\mathbf u^\top\x_i
+
w\log\widetilde g(i\mid H)
\right).
\label{eq:decode}
\end{equation}
The prior $\widetilde g(i\mid H)$ favors items that frequently co-occur with the items in the user's recent history.
We compute it once from the training interactions before model training and keep it fixed. The trust $w\geq0$ controls
its influence. Setting $w=0$ defines the no-prior ablation with a pure dot-product decoder.
Appendix~\ref{app:priors} gives the complete construction.

\subsection{Training objectives and inference}
\label{sec:training}

\textbf{Base flow objective.}
Using the decoder in \eqnref{eq:decode}, we apply next-item classification losses to the behavior source and
the one-step endpoint:
\begin{equation}
\mathcal L_{\mathrm{src}}
=
-\log D_\theta(\x_0,H)[i_{m+1}],
\qquad
\mathcal L_{\mathrm{align}}
=
-\log D_\theta(\widetilde{\x}_1,H)[i_{m+1}].
\label{eq:retrieval-losses}
\end{equation}
Together with the conditional flow-matching loss, the inherited FlowRec objective is
\begin{equation}
\mathcal L_{\mathrm{flow}}
=
\mathcal L_{\mathrm{src}}
+
\lambda_{\mathrm{cfm}}\mathcal L_{\mathrm{CFM}}
+
\lambda_{\mathrm{align}}\mathcal L_{\mathrm{align}}.
\label{eq:flowloss}
\end{equation}
Here, $\mathcal L_{\mathrm{src}}$ makes the history-dependent source predictive of the next item, while
$\mathcal L_{\mathrm{align}}$ applies the same supervision to the one-step endpoint estimate. The weights $\lambda_{\mathrm{cfm}}$ and
$\lambda_{\mathrm{align}}$ are fixed throughout our experiments; their values are given in
Section~\ref{sec:setup}.

\textbf{Code matching.}
The item-level losses can be satisfied through the item-specific vectors $\r_i$ alone, leaving the shared assignments
uninformative for retrieval. We therefore train the predicted endpoint to recover the target item's soft assignments,
encouraging items retrieved in similar contexts to use compatible prototypes. Let
$A^{(s,t)}\in\R^{|\I|\times K}$ collect all item assignments for chain $s$ at depth $t$, and let
$\widetilde{\mathbf g}_H=[\widetilde g(i\mid H)]_{i\in\I}$. Here and below, $\sum_{s,t}$ ranges over all $S$
chains and $T$ depths, with $T=1$ outside the residual-SID setting. We define
\begin{equation}
\begin{aligned}
\boldsymbol\pi_H^{(s,t)}
&=
\operatorname{sg}\!\left(A^{(s,t)\top}\widetilde{\mathbf g}_H\right),
&\qquad
q_H^{(s,t)}
&=
\operatorname{softmax}\!\left(
\widetilde{\x}_1C^{(s,t)\top}
+
w\log\boldsymbol\pi_H^{(s,t)}
\right).
\end{aligned}
\label{eq:code-prediction}
\end{equation}
Here, $\boldsymbol\pi_H^{(s,t)}$ marginalizes the fixed item-space prior into code space. The stop-gradient operator
$\operatorname{sg}$ prevents this computation from updating the assignments. The code-matching loss is
\begin{equation}
\mathcal L_{\mathrm{match}}
=
-\frac{1}{ST}\sum_{s,t}
\left\langle A_{i_{m+1}}^{(s,t)},\log q_H^{(s,t)}\right\rangle.
\label{eq:match}
\end{equation}
Gradients still pass through the target assignment and, through $q_H^{(s,t)}$, to the prototypes and flow model,
allowing these components to co-adapt.

\textbf{Maintaining prototype utilization.}
Because the item-specific channel can support retrieval even when assignments concentrate on only a few prototypes, we
penalize catalog-wide utilization below a target $K_{\mathrm{eff}}$:
\begin{equation}
\bar A^{(s,t)}
=
\frac{1}{|\I|}\sum_{i\in\I}A_i^{(s,t)},
\qquad
\mathcal L_{\mathrm{bal}}
=
\frac{1}{ST}\sum_{s,t}
\left[
\log K_{\mathrm{eff}}
-
\mathcal H\!\left(\bar A^{(s,t)}\right)
\right]_+,
\label{eq:bal}
\end{equation}
where $\mathcal H$ is entropy and $[z]_+=\max(z,0)$. Since
$\exp(\mathcal H(\bar A^{(s,t)}))$ is the effective number of utilized prototypes, the loss vanishes once it reaches
$K_{\mathrm{eff}}$. The loss depends only on the catalog-wide marginal $\bar A^{(s,t)}$, leaving individual assignments
free to be soft or sharp without forcing discretization.

\textbf{Full objective.}
All trainable components are optimized jointly. Given weights $\lambda_{\mathrm{match}},\lambda_{\mathrm{bal}}\geq0$,
\begin{equation}
\mathcal L
=
\mathcal L_{\mathrm{flow}}
+
\lambda_{\mathrm{match}}\mathcal L_{\mathrm{match}}
+
\lambda_{\mathrm{bal}}\mathcal L_{\mathrm{bal}}.
\label{eq:loss}
\end{equation}

\textbf{Inference.}
After training, we materialize the item table $\mX$ from \eqnref{eq:compose}, retaining the learned soft
mixtures without discretization. Given a history $H$, we look up its item embeddings and compute the behavior source
$\x_0$. Starting from $\x_0$, we numerically integrate the learned velocity field as in \eqnref{eq:ode} to
obtain $\hat\x_1$. The decoder in \eqnref{eq:decode} then scores the catalog, and we return the top-$k$ items.
Because the table is materialized once, the prototype-mixture encoder adds no per-query computation relative to a
flat item table.

\section{Experiments}
\label{sec:experiments}

Our experiments address three questions: (\textbf{RQ1}) when does a prototype-mixture item table improve over flat
and hard SID alternatives under different levels of pretrained information; (\textbf{RQ2}) how does the complete
\methodname\ system compare with continuous and discrete generative recommenders in predictive performance and
efficiency; and (\textbf{RQ3}) how do semantic initialization, the co-visitation prior, and target-item frequency
affect the observed gains?

\subsection{Setup}
\label{sec:setup}

\textbf{Datasets and preprocessing.} We evaluate on six public benchmarks: Amazon \emph{Beauty} (2014)~\citep{mcauley2015image,10.1145/2872427.2883037}, Amazon \emph{Industrial \& Scientific} and Amazon \emph{Office} (2023)~\citep{hou2024bridging}, \emph{MovieLens-1M}~\citep{harper2015movielens}, \emph{Yelp-2022}~\citep{yelpdataset}, and \emph{Steam-2018}~\citep{kang2018sasrec}.
All datasets are processed identically following prior work~\citep{geng2022p5,rajput2023tiger,shah2025madrec}:
iterative $5$-core filtering, consecutive-repeat removal, chronological ordering, and leave-one-out splitting, with each user's penultimate and final interactions used for validation and testing.
Appendix~\ref{app:datasets} reports dataset statistics, sources, metadata, and complete preprocessing details.

\textbf{Evaluation protocol.} All models are trained and evaluated on identical item sequences using
full-catalog ranking without candidate sampling. We report HR@$5$~$\uparrow$ and NDCG@$5$~$\uparrow$ as three-seed means scaled by $100$; Appendix~\ref{app:fullresults} additionally reports
$k\in\{10,20\}$. We use early stopping, select the checkpoint with the best validation performance, and evaluate only that checkpoint on the test split.

\textbf{Information regimes.} To separate content information from parameter-sharing benefits, we evaluate three regimes that differ in the pretrained information available. \scen{From scratch} provides none: item representations are initialized randomly
and learned from interactions. \scen{Offline SID} provides the stored codes and posterior responsibilities of a
balanced RQ-GMM (see Appendix~\ref{app:rqgmm}), but not the content embeddings from which they were computed. \scen{PCA-text} provides the full
content embeddings, projected through PCA to $D$ dimensions and kept trainable.

\textbf{Implementation and model selection.}
All controlled item-table variants use the same $D{=}128$ flow backbone, three-item histories, three Euler steps,
and matched non-table hyperparameters. We encode item metadata once with frozen
Qwen3-Embedding-4B~\citep{qwen3embedding}. In \scen{PCA-text}, we initialize the trainable item-specific vectors with standardized $D$-dimensional PCA projections of the content embeddings. \methodname\ zero-initializes its prototypes, making its initial
table identical to the flat baseline. We select configurations by the three-seed validation geometric mean of
HR@5~$\uparrow$ and NDCG@5~$\uparrow$. Appendices~\ref{app:datasets}--\ref{app:rqgmm} give the SID construction and remaining setup details.

\subsection{RQ1: When do prototype mixtures help?}
\label{sec:scenarios}

Our central hypothesis is that continuous item embeddings benefit from shared prototypes when that structure
complements, rather than replaces, an item-specific continuous representation. We test this by holding the flow
backbone and training protocol fixed while varying the item-table parameterization. Because the same table embeds
histories, defines the flow targets, and supports retrieval, ranking differences measure the downstream utility of
the learned item embeddings.

\textbf{Item-table variants.}
The \armname{flat} table uses $\x_i=\r_i$. The \armname{hard} SID table uses
$\x_i=\r_i+\sum_s C^{(s)}[c_i^{(s)}]$, with trainable codewords and item-specific vectors but frozen one-hot
assignments. \methodname\ instead uses the prototype-mixture representation of \eqnref{eq:compose}. All variants
share the same backbone and non-table hyperparameters.

\textbf{Information available to each variant.} Table~\ref{tab:scenarios} compares the item-table variants under three settings. In \scen{from scratch}, the flat and \methodname\ tables are randomly initialized and learn only from interactions. In \scen{offline SID}, the hard table uses the tokenizer's stored codes as fixed assignments, whereas \methodname\ uses its posterior responsibilities as the assignment prior in \eqnref{eq:sidprior}; neither receives the original content embedding. In \scen{PCA-text}, every variant initializes its item-specific vectors from the same PCA projection. The hard table uses the tokenizer's fixed codes, whereas \methodname\ initializes its prototypes to zero, making its initial table identical to the flat table.

\begin{table}[t]
\centering\small
\caption{\textbf{Item tables across three information regimes.} Each cell reports test HR@5~$\uparrow$ (top)
and change from the dataset's scratch \armname{flat} anchor (bottom; \emph{ref.}). The final $\Delta$ compares
\methodname\ with \armname{flat} (scratch), \armname{hard} (offline SID), or \armname{flat} (PCA-text).
\armname{hard} uses the same tokenizer; all variants have matched backbones and non-table hyperparameters.}
\label{tab:scenarios}
\resizebox{\linewidth}{!}{\providecommand{\pending}[1]{\textcolor{red}{#1}}
\providecommand{\scen}[1]{\textsc{#1}}
\providecommand{\armname}[1]{\textsc{#1}}

\begin{tabular}{@{}l ccc ccc cccc@{}}
\toprule
 & \multicolumn{3}{c}{\textbf{\scen{from scratch}}} & \multicolumn{3}{c}{\textbf{\scen{offline SID}}} & \multicolumn{4}{c}{\textbf{\scen{PCA-text}}} \\
\cmidrule(lr){2-4}\cmidrule(lr){5-7}\cmidrule(lr){8-11}
dataset & \armname{flat} & \methodname & \emph{vs.} \armname{flat} & \armname{hard} & \methodname & \emph{vs.} \armname{hard} & \armname{flat} & \armname{hard} & \methodname & \emph{vs.} \armname{flat} \\
\midrule
Beauty & \makecell{$5.77$\\{\scriptsize\emph{ref.}}} & \makecell{$5.89$\\{\scriptsize\textcolor{green!55!black}{$+2.08\%$}}} & \textcolor{green!55!black}{$+2.08\%$} & \makecell{$5.80$\\{\scriptsize\textcolor{green!55!black}{$+0.52\%$}}} & \makecell{$6.32$\\{\scriptsize\textcolor{green!55!black}{$+9.53\%$}}} & \textcolor{green!55!black}{$+8.97\%$} & \makecell{$6.69$\\{\scriptsize\textcolor{green!55!black}{$+15.94\%$}}} & \makecell{$6.43$\\{\scriptsize\textcolor{green!55!black}{$+11.44\%$}}} & \makecell{$6.79$\\{\scriptsize\textcolor{green!55!black}{$+17.68\%$}}} & \textcolor{green!55!black}{$+1.49\%$} \\
Industrial & \makecell{$2.75$\\{\scriptsize\emph{ref.}}} & \makecell{$2.88$\\{\scriptsize\textcolor{green!55!black}{$+4.73\%$}}} & \textcolor{green!55!black}{$+4.73\%$} & \makecell{$3.02$\\{\scriptsize\textcolor{green!55!black}{$+9.82\%$}}} & \makecell{$3.47$\\{\scriptsize\textcolor{green!55!black}{$+26.18\%$}}} & \textcolor{green!55!black}{$+14.90\%$} & \makecell{$3.81$\\{\scriptsize\textcolor{green!55!black}{$+38.55\%$}}} & \makecell{$3.52$\\{\scriptsize\textcolor{green!55!black}{$+28.00\%$}}} & \makecell{$3.79$\\{\scriptsize\textcolor{green!55!black}{$+37.82\%$}}} & \textcolor{red!70!black}{$-0.52\%$} \\
Steam & \makecell{$8.29$\\{\scriptsize\emph{ref.}}} & \makecell{$8.35$\\{\scriptsize\textcolor{green!55!black}{$+0.72\%$}}} & \textcolor{green!55!black}{$+0.72\%$} & \makecell{$8.32$\\{\scriptsize\textcolor{green!55!black}{$+0.36\%$}}} & \makecell{$8.69$\\{\scriptsize\textcolor{green!55!black}{$+4.83\%$}}} & \textcolor{green!55!black}{$+4.45\%$} & \makecell{$8.13$\\{\scriptsize\textcolor{red!70!black}{$-1.93\%$}}} & \makecell{$8.19$\\{\scriptsize\textcolor{red!70!black}{$-1.21\%$}}} & \makecell{$8.30$\\{\scriptsize\textcolor{green!55!black}{$+0.12\%$}}} & \textcolor{green!55!black}{$+2.09\%$} \\
\bottomrule
\end{tabular}
}
\end{table}

\textbf{Prototype sharing helps without pretrained content.}
In \scen{from scratch}, \methodname\ improves HR@5~$\uparrow$ over the flat table by
\QoneVsFlatBeauty\ on Beauty, \QoneVsFlatIndus\ on Industrial, and \QoneVsFlatSteam\ on Steam. Thus, shared
prototypes learn useful cross-item structure from interactions alone, producing item embeddings that are more effective for next-item ranking than independently parameterized rows.

\textbf{An offline SID is more useful as a prior than as a fixed assignment.}
The hard SID table exceeds the scratch anchor on all three datasets, confirming that the tokenizer carries useful
information. Given the same tokenizer, \methodname\ improves HR@5~$\uparrow$ over the hard table by
\QtwoaRangeLo--\QtwoaRangeHi, showing that differentiable assignments use this information more effectively than
frozen one-hot assignments.

\textbf{Prototype sharing complements the full content embedding.}
In \scen{PCA-text}, content seeding already strengthens the flat table on the Amazon datasets. Starting from the
same embeddings, \methodname\ improves it by \QtwobVsFlatBeauty\ on Beauty and \QtwobVsFlatSteam\ on Steam while
essentially preserving its Industrial result (\QtwobVsFlatIndus). In contrast, the hard SID table changes
HR@5~$\uparrow$ by \HardVsFlatPcaBeauty\ and \HardVsFlatPcaIndus\ relative to the seeded flat table on the Amazon
datasets. Thus, prototype mixtures add cross-item sharing while retaining the unquantized initialization improvement.

Across all three regimes, prototype mixtures improve how continuous item embeddings use the information available by introducing cross-item pooling without quantizing their representations.

\subsection{RQ2: Comparison with generative recommenders}
\label{sec:compare}

We compare \methodname\ with continuous FlowRec~\citep{li2025preferencetrajectorymodelingflow}, autoregressive
TIGER~\citep{rajput2023tiger}, and masked-diffusion MaskGR~\citep{shah2025madrec} in all three information regimes,
selecting one configuration per model and dataset by the validation geometric mean of HR@5~$\uparrow$ and
NDCG@5~$\uparrow$. \methodname\ and FlowRec use three-item histories; TIGER and MaskGR retain their released
$120$-token limit and decode one top-$20$ list with beam search. Appendix~\ref{app:impl} specifies the tokenizers,
initializations, resulting item-level history lengths, and decoding protocol.

Table~\ref{tab:main-summary} reports the complete-system comparison at $k{=}5$, with results at $k{=}10$ and
$20$ in Appendix~\ref{app:fullresults}. \methodname\ achieves the highest HR@5~$\uparrow$ on all six datasets and
the highest or tied-for-highest NDCG@5~$\uparrow$ on all six. Averaged across datasets, it improves over the strongest
evaluated baseline for each dataset and metric by \AvgHRfive\ in HR@5~$\uparrow$ and \AvgNDCGfive\ in
NDCG@5~$\uparrow$. For \methodname, validation selects \scen{PCA-text} on five of the six datasets and
\scen{offline SID} on Steam, suggesting that direct content initialization is generally preferable when the
original embeddings are available.

\begin{table}[t]
\centering\small
\caption{\textbf{Main results}, HR@5~$\uparrow$ and NDCG@5~$\uparrow$
(mean$\pm$std, $\times100$). For each model--dataset pair, both metrics come from one run selected by their
validation geometric mean; $^{\text{s}}$ marks scratch, $^{\text{i}}$ offline SID, and $^{\text{p}}$ PCA-text.
\textbf{Bold}/\underline{underline} mark best/second; $\Delta$ compares \methodname\ with the best baseline.}
\label{tab:main-summary}
\resizebox{\linewidth}{!}{\providecommand{\pending}[1]{\textcolor{red}{#1}}

\setlength{\tabcolsep}{2.7pt}
\renewcommand{\cellalign}{tc}
\begin{tabular}{@{}l cc cc cc cc cc cc@{}}
\toprule
 & \multicolumn{2}{c}{\textbf{Beauty}} & \multicolumn{2}{c}{\textbf{Indus.}} & \multicolumn{2}{c}{\textbf{Steam}} & \multicolumn{2}{c}{\textbf{ML-1M}} & \multicolumn{2}{c}{\textbf{Yelp}} & \multicolumn{2}{c}{\textbf{Office}} \\
\cmidrule(lr){2-3}\cmidrule(lr){4-5}\cmidrule(lr){6-7}\cmidrule(lr){8-9}\cmidrule(lr){10-11}\cmidrule(lr){12-13}
 & HR@5 $\uparrow$ & NDCG@5 $\uparrow$ & HR@5 $\uparrow$ & NDCG@5 $\uparrow$ & HR@5 $\uparrow$ & NDCG@5 $\uparrow$ & HR@5 $\uparrow$ & NDCG@5 $\uparrow$ & HR@5 $\uparrow$ & NDCG@5 $\uparrow$ & HR@5 $\uparrow$ & NDCG@5 $\uparrow$ \\
\midrule
FlowRec & \makecell{\underline{6.27}$^{\text{p}}$\\[-3pt]{\tiny$\pm$0.18}} & \makecell{\underline{4.03}\\[-3pt]{\tiny$\pm$0.03}} & \makecell{\underline{3.62}$^{\text{p}}$\\[-3pt]{\tiny$\pm$0.02}} & \makecell{2.04\\[-3pt]{\tiny$\pm$0.03}} & \makecell{8.28$^{\text{s}}$\\[-3pt]{\tiny$\pm$0.33}} & \makecell{\underline{5.37}\\[-3pt]{\tiny$\pm$0.30}} & \makecell{15.36$^{\text{p}}$\\[-3pt]{\tiny$\pm$0.20}} & \makecell{10.14\\[-3pt]{\tiny$\pm$0.12}} & \makecell{2.98$^{\text{p}}$\\[-3pt]{\tiny$\pm$0.05}} & \makecell{1.86\\[-3pt]{\tiny$\pm$0.03}} & \makecell{\underline{2.90}$^{\text{p}}$\\[-3pt]{\tiny$\pm$0.02}} & \makecell{\underline{1.67}\\[-3pt]{\tiny$\pm$0.02}} \\
TIGER & \makecell{3.51$^{\text{s}}$\\[-3pt]{\tiny$\pm$0.07}} & \makecell{2.54\\[-3pt]{\tiny$\pm$0.05}} & \makecell{2.05$^{\text{p}}$\\[-3pt]{\tiny$\pm$0.03}} & \makecell{1.34\\[-3pt]{\tiny$\pm$0.02}} & \makecell{8.43$^{\text{i}}$\\[-3pt]{\tiny$\pm$0.10}} & \makecell{5.01\\[-3pt]{\tiny$\pm$0.03}} & \makecell{12.59$^{\text{s}}$\\[-3pt]{\tiny$\pm$0.40}} & \makecell{8.47\\[-3pt]{\tiny$\pm$0.28}} & \makecell{1.29$^{\text{p}}$\\[-3pt]{\tiny$\pm$0.07}} & \makecell{0.80\\[-3pt]{\tiny$\pm$0.04}} & \makecell{1.52$^{\text{p}}$\\[-3pt]{\tiny$\pm$0.01}} & \makecell{0.98\\[-3pt]{\tiny$\pm$0.01}} \\
MaskGR & \makecell{5.06$^{\text{i}}$\\[-3pt]{\tiny$\pm$0.13}} & \makecell{3.28\\[-3pt]{\tiny$\pm$0.12}} & \makecell{3.46$^{\text{p}}$\\[-3pt]{\tiny$\pm$0.03}} & \makecell{\underline{2.24}\\[-3pt]{\tiny$\pm$0.03}} & \makecell{\underline{8.64}$^{\text{p}}$\\[-3pt]{\tiny$\pm$0.00}} & \makecell{\textbf{5.57}\\[-3pt]{\tiny$\pm$0.01}} & \makecell{\underline{15.64}$^{\text{s}}$\\[-3pt]{\tiny$\pm$0.40}} & \makecell{\underline{10.23}\\[-3pt]{\tiny$\pm$0.28}} & \makecell{\underline{2.99}$^{\text{s}}$\\[-3pt]{\tiny$\pm$0.04}} & \makecell{\underline{1.87}\\[-3pt]{\tiny$\pm$0.03}} & \makecell{2.29$^{\text{s}}$\\[-3pt]{\tiny$\pm$0.02}} & \makecell{1.53\\[-3pt]{\tiny$\pm$0.01}} \\
\textbf{\methodname} & \makecell{\textbf{6.79}$^{\text{p}}$\\[-3pt]{\tiny$\pm$0.06}} & \makecell{\textbf{4.73}\\[-3pt]{\tiny$\pm$0.04}} & \makecell{\textbf{3.79}$^{\text{p}}$\\[-3pt]{\tiny$\pm$0.01}} & \makecell{\textbf{2.34}\\[-3pt]{\tiny$\pm$0.03}} & \makecell{\textbf{8.69}$^{\text{i}}$\\[-3pt]{\tiny$\pm$0.29}} & \makecell{\textbf{5.57}\\[-3pt]{\tiny$\pm$0.19}} & \makecell{\textbf{17.25}$^{\text{p}}$\\[-3pt]{\tiny$\pm$0.13}} & \makecell{\textbf{11.71}\\[-3pt]{\tiny$\pm$0.11}} & \makecell{\textbf{3.18}$^{\text{p}}$\\[-3pt]{\tiny$\pm$0.02}} & \makecell{\textbf{2.03}\\[-3pt]{\tiny$\pm$0.02}} & \makecell{\textbf{3.22}$^{\text{p}}$\\[-3pt]{\tiny$\pm$0.01}} & \makecell{\textbf{2.28}\\[-3pt]{\tiny$\pm$0.01}} \\
$\Delta$ (\%) & \textcolor{green!55!black}{$+8.3\%$} & \textcolor{green!55!black}{$+17.4\%$} & \textcolor{green!55!black}{$+4.7\%$} & \textcolor{green!55!black}{$+4.5\%$} & \textcolor{green!55!black}{$+0.6\%$} & $0.0\%$ & \textcolor{green!55!black}{$+10.3\%$} & \textcolor{green!55!black}{$+14.5\%$} & \textcolor{green!55!black}{$+6.4\%$} & \textcolor{green!55!black}{$+8.6\%$} & \textcolor{green!55!black}{$+11.0\%$} & \textcolor{green!55!black}{$+36.5\%$} \\
\bottomrule
\end{tabular}
}
\end{table}

\textbf{Inference efficiency.}
At inference, \methodname\ materializes the composed item embeddings once, so its soft assignments add no per-query
computation. Table~\ref{tab:cost-summary} shows that \methodname\ consistently requires fewer MFLOP/query~$\downarrow$
and achieves higher throughput~$\uparrow$ than both discrete baselines across the five profiled datasets. The
advantage is largest when the baselines generate multi-code SIDs and narrows for atomic-token configurations. The
parameter advantage depends on catalog size: on Yelp, the continuous item table makes \methodname\ slightly larger
than TIGER. Appendix~\ref{app:impl} documents the measurement protocol, and Table~\ref{tab:cost} reports the full per-dataset measurements.

\begin{table}[t]
\centering\small
\caption{\textbf{Inference efficiency} across five datasets
(min--max from unrounded values). Hot-parameter ratio~$\uparrow$ and MFLOP/query ratio~$\uparrow$
are baseline/\methodname; throughput ratio~$\uparrow$ is \methodname/baseline.}
\label{tab:cost-summary}
\setlength{\tabcolsep}{8pt}
\begin{tabular}{@{}lccc@{}}
\toprule
baseline & hot-parameter ratio $\uparrow$ & MFLOP/query ratio $\uparrow$ & throughput ratio $\uparrow$ \\
\midrule
TIGER  & $0.9$--$10.0\times$ & $23$--$265\times$       & $1.9$--$7.1\times$ \\
MaskGR & $1.4$--$5.4\times$  & $3.7$--$2{,}654\times$ & $1.4$--$161\times$ \\
\bottomrule
\end{tabular}
\end{table}

\subsection{RQ3: Robustness and diagnostics}
\label{sec:diagnostics}

\textbf{Motivating controls.}
Appendix~\ref{app:diagnostics} compares hard semantic assignments with random assignments matched in depth, codebook size, and codeword vectors. Without content initialization, semantic assignments win on all three datasets, confirming that they carry useful structure beyond parameter sharing. With PCA initialization, they underperform the flat and random controls on both Amazon datasets, showing that this structure can become restrictive when frozen.

\textbf{Complementarity with the retrieval prior.}
Without the co-visitation prior, prototype sharing is beneficial in all regimes on both Amazon datasets
and retains a \NoPriorSidSteam\ HR@5~$\uparrow$ advantage over hard SID on Steam
(Table~\ref{tab:noprior}), confirming that prototype sharing and the prior are complementary.

\textbf{How do gains vary with item frequency?}
In the \scen{from scratch} regime, the frequency-stratified results are mixed. However, when pretrained tokenizer or content information is available, relative gains are larger for rare targets than for frequent ones in $11$ of $12$ dataset--comparison--metric combinations for HR@5~$\uparrow$ and NDCG@10~$\uparrow$, consistent with transfer to less-observed items (Table~\ref{tab:frequency-summary}).

\section{Related Work}
\label{sec:related}

\textbf{Discrete generative recommenders and semantic IDs.}
The discrete branch commonly replaces an atomic item identifier with a \emph{semantic ID} (SID) by quantizing a pretrained content embedding~\citep{grid}. This is generally done through residual schemes that cluster either directly in the embedding space (RQ-KMeans and
RQ-GMM~\citep{deng2025onerecunifyingretrieverank,rqgmm2026}) or in a learned latent space
(RQ-VAE~\citep{Oord2017,LeeKKCH22}). A recommender then predicts the resulting tuple autoregressively, as in
TIGER~\citep{rajput2023tiger}, OneRec~\citep{deng2025onerecunifyingretrieverank,zhou2025onerectechnicalreport}, and
MiniOneRec~\citep{kong2025minionerecopensourceframeworkscaling}, or through masked discrete diffusion, as in
MaskGR~\citep{shah2025madrec}. SIDs provide two distinct benefits. The first, \emph{parameter sharing}, arises because reusing codewords ties parameters across items and pools supervision from their interactions. The second,
\emph{semantic transfer}, comes from deriving the assignments from pretrained content embeddings, which places
semantically related items under shared parameters. Together, these benefits improve generalization~\citep{grid,singh2024better}.
However, an SID exposes only hard code assignments and preserves a partition of the content space rather than its full continuous geometry.

\textbf{Diffusion and flow recommenders.}
Continuous generative recommenders replace discrete token prediction with generation in a continuous space.
DiffRec~\citep{10.1145/3539618.3591663} denoises a user's interaction vector, whereas DreamRec~\citep{yang2023generate}
and DiffuRec~\citep{10.1145/3631116} diffuse a target item embedding for next-item prediction. FMRec~\citep{10.24963/ijcai.2025/346}
and FlowRec~\citep{li2025preferencetrajectorymodelingflow} use flow matching~\citep{lipman_flow_2023} to
transport a source distribution to the target embedding, with FlowRec conditioning the source on the user's history. These methods improve the generative dynamics while retaining a flat item table, thereby avoiding
discretization into code sequences but forgoing the cross-item parameter sharing induced by shared SID codewords. \methodname\ retains
continuous generation while introducing that sharing through co-adaptive mixtures of prototypes.

\section{Conclusion}
\label{sec:conclusion}

We presented \methodname, a flow-based recommender whose item table combines differentiable assignments over shared codebooks with an item-specific vector, all trained jointly with the recommendation objective.
The shared codewords introduce explicit cross-item parameter coupling, while the item-specific vectors retain unrestricted per-item degrees of freedom.
Across six public benchmarks, \methodname\ improves average HR@5~$\uparrow$ and NDCG@5~$\uparrow$ relative to the strongest evaluated baselines, while providing higher throughput~$\uparrow$ than the evaluated discrete baselines.

\bibliographystyle{assets/plainnat}
\bibliography{refs,fm}

\clearpage
\newpage
\beginappendix
\setcounter{topnumber}{3}
\setcounter{totalnumber}{5}
\renewcommand{\topfraction}{0.9}
\renewcommand{\bottomfraction}{0.8}
\renewcommand{\textfraction}{0.07}
\renewcommand{\floatpagefraction}{0.8}

\section{Dataset details}
\label{app:datasets}

\textbf{Sources.} Most datasets are taken as RecBole atomic files~\citep{zhao2021recbole} from the RecSysDatasets\footnote{\url{https://github.com/RUCAIBox/RecSysDatasets}} distribution. Amazon's Industrial \&
Scientific and Office are taken directly from the raw Amazon Reviews 2023 website\footnote{\url{https://amazon-reviews-2023.github.io/}}. Regardless of the source, all datasets are reduced to \texttt{(user, item, timestamp)} triples and then processed through the same code path.

\begin{table}[h]\centering\small
\caption{\textbf{Datasets after iterative 5-core filtering}. \emph{Avg.\ length} is interactions per user, \emph{Median inter./item} is the median
number of interactions a catalog item receives, and sparsity is $1-\#$interactions$/(\#$users$\cdot\#$items$)$. \emph{Eff.\ rank} is $\exp H(p)$ over the normalized eigenvalue spectrum of the content embedding, out of $2560$, and \emph{Var.\ kept} is the share of that variance surviving the projection to $D$.}
\label{tab:datasets}
\resizebox{\linewidth}{!}{\begin{tabular}{@{}l rrrrrr @{\hspace{8pt}}|@{\hspace{8pt}} rr@{}}
\toprule
 & \multicolumn{6}{c}{\textbf{Catalog}} & \multicolumn{2}{c}{\textbf{Content embedding}} \\
\cmidrule(lr){2-7}\cmidrule(l){8-9}
Dataset & \#Users & \#Items & \#Interactions & Avg.\ length & Median inter./item & Sparsity (\%) & Eff.\ rank & Var.\ kept (\%) \\
\midrule
Amazon Beauty                   & 22,363 & 12,101 & 198,502 & 8.9 & 9 & 99.927 & 219 & 73.2 \\
Amazon Industrial \& Scientific & 51,322 & 25,923 & 417,422 & 8.1 & 9 & 99.969 & 240 & 73.5 \\
Amazon Office                   & 223,996 & 77,655 & 1,810,031 & 8.1 & 11 & 99.990 & 213 & 75.0 \\
MovieLens-1M                    & 6,040 & 3,416 & 999,611 & 165.5 & 146 & 95.155 & 326 & 66.0 \\
Yelp-2022                       & 272,547 & 111,260 & 4,185,885 & 15.4 & 14 & 99.986 & 279 & 71.3 \\
Steam                           & 25,256 & 4,087 & 325,320 & 12.9 & 18 & 99.685 & 304 & 67.6 \\
\bottomrule
\end{tabular}
}
\end{table}

\textbf{Preprocessing.} Every dataset follows the same procedure. First, we apply an \emph{iterative} $k$-core filter ($k=5$). Removing a low-count user can push an item below the threshold and vice versa, so we alternately drop
users with fewer than $k$ interactions and items with fewer than $k$ interactions until a pass removes nothing.
Second, each user's interactions are sorted chronologically by timestamp. Third, we collapse consecutive repeats of
the same item and then \emph{re-run} the $k$-core filter, iterating both to a fixed point. Iterating guarantees every
surviving user has at least $k$ distinct items and therefore appears in all three splits. In our experiments,
datasets require between $1$ and $4$ collapse-and-re-core rounds. We then assign dense user and item IDs. Finally, each user's sequence $(i_1,\dots,i_n)$ is split using a leave-one-out scheme: the test
instance is the full sequence with target $i_n$, the validation instance is $(i_1,\dots,i_{n-1})$ with target
$i_{n-1}$, and training uses $(i_1,\dots,i_{n-2})$.
All model families consume identical item sequences.

\textbf{Text embedding.} Every dataset has a deterministic per-item text template, filled from the item metadata
and left empty where a field is missing: Amazon uses \texttt{Title; Brand; Categories; Price}, following
TIGER~\citep{rajput2023tiger}; MovieLens uses
\texttt{Title; Release year; Genres} plus the most frequent community tags; Yelp uses \texttt{Name; City, State;
Categories}; and Steam uses \texttt{Title; Developer; Publisher; Genres; Tags}. Each string is encoded once with a
frozen Qwen3-Embedding-4B~\citep{qwen3embedding} using last-token pooling, giving $\mathbf{q}_i\in\R^{2560}$.

\textbf{SID pretraining.} The offline SIDs are produced by the balanced RQ-GMM of Appendix~\ref{app:rqgmm}. We fit $4$ levels
with $K\in\{256,512\}$, plus a fifth de-duplication digit that makes identifiers unique. The hard RQ-SID baseline consumes the resulting codes directly, while \methodname\ consumes the per-level posterior responsibilities as the prior in \eqnref{eq:sidprior}.

\textbf{Evaluation.} Evaluation ranks the target against the \emph{full} catalog, with no candidate sampling and
no filtering of previously seen items, so every model scores all $|\I|$ items for every test user.
We report hit rate and normalized discounted cumulative gain.
Hit rate (HR@$k$~$\uparrow$) is the fraction of users whose held-out item appears among the $k$ highest-ranked
items. Normalized discounted cumulative gain (NDCG@$k$~$\uparrow$) additionally rewards placing that item higher
within the top $k$. We report $k\in\{5,10,20\}$.
The two generative baselines decode differently from \methodname\ but are scored on the same
ranked list: TIGER produces its top-$20$ by autoregressive beam search over code positions, and MaskGR by
beam search over the positions it unmasks. Every number is a mean over three seeds, taken from the checkpoint with the best validation performance in each run.

\FloatBarrier
\section{Implementation details}
\label{app:impl}

\textbf{Architecture and optimization.}
The history encoder is a bidirectional Transformer with rotary positions, RMSNorm, and pre-norm blocks ($4$
layers, $4$ heads, and hidden dimension $D{=}128$), reading the $m{=}3$ most recent items as one token each. The
time code $\varphi(t)$ is produced by passing a $128$-dimensional sinusoidal embedding through a two-layer MLP with
SiLU activations. The velocity field is a second two-layer SiLU MLP with hidden width $2D{=}256$, applied to the
concatenation of the modulated state and $\varphi(t)$. We use neither normalization nor FiLM conditioning:
\begin{equation}
v_\theta(\x_t,t)
=
\operatorname{MLP}\!\left([g(\x_t);\varphi(t)]\right),
\qquad
g(\x_t)
=
\begin{cases}
\boldsymbol\lambda\odot\x_t,
& \lambda_j\overset{\mathrm{iid}}{\sim}\mathcal N(\delta,\delta^2)
\quad\text{during training},\\
\delta\x_t,
& \text{during evaluation},
\end{cases}
\label{eq:field}
\end{equation}
where $\delta=10^{-3}$. The modulation vector is redrawn for every call during training and replaced by its mean
during evaluation. Item features have dimension $D_f{=}32$. The assignment head in \eqnref{eq:assign} uses a separate
affine input projection and output projection for each codebook. With $T=1$, the $S$ codebook tokens are processed
jointly by a single-layer, single-head bidirectional self-attention block. With a residual SID, each of the $S$
chains contains $T$ tokens and is processed independently with causal attention following the tokenizer's level
order; the depth-$t$ prior is shared across chains. We optimize
with AdamW at learning rate $10^{-3}$ and no weight decay, use dropout $0.3$ on item embeddings and $0.5$ on hidden
states, and stop early with patience $50$. Inference uses $N{=}3$ Euler steps.

\textbf{Hyperparameter search.}
We fix $\lambda_{\mathrm{cfm}}{=}10$ and $\lambda_{\mathrm{align}}{=}1$, use Optuna's TPE
sampler~\citep{bergstra2011tpe,akiba2019optuna}, and select configurations by the three-seed validation geometric
mean of HR@5~$\uparrow$ and NDCG@5~$\uparrow$.
We independently sample $\lambda_{\mathrm{match}}$ log-uniformly from $[10^{-4},4]$,
$\lambda_{\mathrm{bal}}$ log-uniformly from $[10^{-4},1]$, and $K_{\mathrm{eff}}/K$ log-uniformly from
$[0.02,1]$. We search $S\in\{1,2,4\}$ in all settings; $T=1$ by default, whereas
$T\in\{1,2,3\}$ is selected jointly with $S$ in the offline-SID setting.

\textbf{PCA-text initialization.}
Within each run, we fit PCA on the catalog's $2560$-dimensional content embeddings and project them to $D{=}128$.
We then standardize each projected coordinate across the catalog to zero mean and unit variance, preventing any
principal direction from dominating the initialization. For the continuous models, these standardized vectors
initialize the trainable item-specific vectors $\r_i$. \methodname\ initializes every prototype to zero, so
\eqnref{eq:compose} gives the same initial item table as the flat PCA-initialized model. For TIGER and MaskGR, we
apply the same fitted PCA projection and coordinatewise standardization to the RQ centroids and use the resulting
vectors to initialize each level's codeword embeddings. The de-duplication level has no centroid and remains randomly
initialized.

\textbf{Baseline implementations.}
We use MaskGR's official implementation\footnote{\url{https://github.com/snap-research/MaskGR}}. Because its
default hyperparameters differ from those reported in its paper, we train both configurations and select between them
using the validation criterion from Section~\ref{sec:setup}.
For TIGER, we use the implementation\footnote{\url{https://github.com/snap-research/GRID}} provided in GRID~\citep{grid}. We supply each with our own SIDs and interaction splits and otherwise keep the published training code untouched.
We could not find an official implementation for FlowRec, so we use our own implementation following \citet{li2025preferencetrajectorymodelingflow} for the flow objective.
For fair comparison, we use the same encoder, velocity field, and decoder as \methodname.
We fix $m{=}3$ for \methodname\ and FlowRec and do not tune it. TIGER and MaskGR use
their released implementations' maximum input length of $120$ tokens. Because an item occupies one token per SID
level, including the de-duplication level, the corresponding maximum number of history items varies with the
validation-selected SID depth.

\textbf{Generative baselines across the three scenarios.} TIGER and MaskGR consume items as tokens, so a
scenario is a choice of tokenizer and of how its token table is initialized. \scen{Offline SID} uses the RQ-GMM
tuple. \scen{From scratch} uses an atomic tokenizer, one level assigning item $i$ the token $i$, so the token
table is $|\I|$ rows learned from interactions alone. \scen{PCA-text} uses the RQ-GMM tuple with the PCA-initialized
codeword rows described above.

\textbf{Measuring inference cost.} All five dataset columns of Table~\ref{tab:cost} describe the same query: every
model is given the $m{=}3$ most recent items, which for a token-based baseline means an input sequence of length
$3n_h$, where $n_h$ is the number of tokens per item. \emph{Hot params} are the parameters a served model must hold, determined by which modules actually fire during
an inference batch rather than by the checkpoint size: \methodname\ materializes its item table $\x_i$ once into
$[|\I|,D]$ and discards the codebooks, assignment head, and item features, so it serves at the size of a flat
table, whereas a generative decoder has nothing to fold away.
\emph{MFLOP/query} is counted with PyTorch's flop formulas. \emph{Queries/s} is wall-clock throughput on one
H200 with a batch size of $128$ in FP32, measured the same way for every model. We run three independent processes per cell; each
contributes one steady-state estimate after warm-up, and we report the median of the three. Within a process we time the decode
alone, excluding import and checkpoint load, over $100$ batches for the baselines and a $12$\,s window for
\methodname.

\textbf{Measuring training cost.} Table~\ref{tab:cost} also reports training cost. Optimizer \emph{steps} come from the checkpoint that produced the
reported result; the \emph{rate} is measured separately from the elapsed-time difference between training-only runs stopped at $N_1$ and $N_2$ steps, excluding validation. GPU-hours are estimated by dividing the selected checkpoint's optimizer-step count by this rate. All cells
use one H200 with identical dataloader-worker and CPU allocations, and we report \emph{examples/s} to account for differences in effective batch size. GPU-hours are implementation-bound: for TIGER and MaskGR, we use the unmodified implementations.

\begin{table}[tbp]
\centering\small
\caption{\textbf{Inference and training cost} at matched input and output: top-$20$ lists from $3$ history items, with TIGER and MaskGR using beam search with width $20$. \emph{Hot params} are those a served model must hold,
split as \emph{table}~$+$~\emph{backbone} (every backbone is constant across
datasets). Subscripts give each baseline relative to \methodname\ and are computed from unrounded measurements. Throughput is measured on one H200 with a batch size of $128$ in FP32.
\emph{Train GPU-h} is the estimated cost of training the reported model, computed by dividing the optimizer-step count at the selected
checkpoint by a separately measured training rate.}
\label{tab:cost}
\providecommand{\pending}[1]{\textcolor{red}{#1}}
\footnotesize\setlength{\tabcolsep}{2.5pt}
\begin{tabular}{@{}lr@{\,}lr@{\,}lr@{\,}lr@{\,}lr@{\,}l@{}}
\toprule
 & \multicolumn{2}{c}{\textbf{ML-1M}} & \multicolumn{2}{c}{\textbf{Steam}} & \multicolumn{2}{c}{\textbf{Beauty}} & \multicolumn{2}{c}{\textbf{Indus.}} & \multicolumn{2}{c}{\textbf{Yelp}} \\
$|\I|$ & \multicolumn{2}{c}{\scriptsize 3.4k} & \multicolumn{2}{c}{\scriptsize 4.1k} & \multicolumn{2}{c}{\scriptsize 12k} & \multicolumn{2}{c}{\scriptsize 25k} & \multicolumn{2}{c}{\scriptsize 111k} \\
\midrule
\multicolumn{11}{@{}l}{\itshape Hot params (M) $\downarrow$: table $+$ backbone} \\
\textbf{\methodname} & $0.4{+}0.9$ &  & $0.5{+}0.9$ &  & $1.5{+}0.9$ &  & $3.3{+}0.9$ &  & $14.2{+}0.9$ &  \\
TIGER & $0.9{+}12.9$ & {\tiny\color{black!55}$10\times$} & $0.3{+}12.9$ & {\tiny\color{black!55}$9.1\times$} & $3.1{+}12.9$ & {\tiny\color{black!55}$6.5\times$} & $0.3{+}12.9$ & {\tiny\color{black!55}$3.1\times$} & $0.5{+}12.9$ & {\tiny\color{black!55}$0.9\times$} \\
MaskGR & $0.4{+}6.8$ & {\tiny\color{black!55}$5.4\times$} & $0.3{+}6.8$ & {\tiny\color{black!55}$5.0\times$} & $0.2{+}6.8$ & {\tiny\color{black!55}$2.8\times$} & $0.3{+}6.8$ & {\tiny\color{black!55}$1.7\times$} & $14.2{+}6.8$ & {\tiny\color{black!55}$1.4\times$} \\
\midrule
\multicolumn{11}{@{}l}{\itshape MFLOP / query $\downarrow$} \\
\textbf{\methodname} & $6.4$ &  & $6.6$ &  & $8.6$ &  & $12.2$ &  & $34.0$ &  \\
TIGER & $195$ & {\tiny\color{black!55}$30\times$} & $1{,}742$ & {\tiny\color{black!55}$265\times$} & $199$ & {\tiny\color{black!55}$23\times$} & $1{,}742$ & {\tiny\color{black!55}$143\times$} & $1{,}747$ & {\tiny\color{black!55}$51\times$} \\
MaskGR & $43.6$ & {\tiny\color{black!55}$6.8\times$} & $17{,}435$ & {\tiny\color{black!55}$2{,}654\times$} & $17{,}037$ & {\tiny\color{black!55}$1{,}976\times$} & $10{,}399$ & {\tiny\color{black!55}$855\times$} & $126$ & {\tiny\color{black!55}$3.7\times$} \\
\midrule
\multicolumn{11}{@{}l}{\itshape Queries/s $\uparrow$} \\
\textbf{\methodname} & $26{,}934$ &  & $28{,}752$ &  & $29{,}240$ &  & $27{,}864$ &  & $7{,}688$ &  \\
TIGER & $5{,}568$ & {\tiny\color{black!55}$1/4.8$} & $4{,}054$ & {\tiny\color{black!55}$1/7.1$} & $7{,}242$ & {\tiny\color{black!55}$1/4.0$} & $4{,}121$ & {\tiny\color{black!55}$1/6.8$} & $4{,}028$ & {\tiny\color{black!55}$1/1.9$} \\
MaskGR & $5{,}634$ & {\tiny\color{black!55}$1/4.8$} & $182$ & {\tiny\color{black!55}$1/158$} & $182$ & {\tiny\color{black!55}$1/161$} & $242$ & {\tiny\color{black!55}$1/115$} & $5{,}656$ & {\tiny\color{black!55}$1/1.4$} \\
\midrule
\multicolumn{11}{@{}l}{\itshape Train GPU-h $\downarrow$} \\
\textbf{\methodname} & $1.46$ &  & $0.21$ &  & $0.06$ &  & $0.08$ &  & $1.99$ &  \\
TIGER & $2.01$ & {\tiny\color{black!55}$1.4\times$} & $0.20$ & {\tiny\color{black!55}$1.0\times$} & $0.53$ & {\tiny\color{black!55}$8.9\times$} & $1.08$ & {\tiny\color{black!55}$13\times$} & $2.21$ & {\tiny\color{black!55}$1.1\times$} \\
MaskGR & $2.06$ & {\tiny\color{black!55}$1.4\times$} & $10$ & {\tiny\color{black!55}$49\times$} & $17$ & {\tiny\color{black!55}$286\times$} & $47$ & {\tiny\color{black!55}$560\times$} & $7.59$ & {\tiny\color{black!55}$3.8\times$} \\
\midrule
\multicolumn{11}{@{}l}{\itshape Train examples/s $\uparrow$} \\
\textbf{\methodname} & $40{,}623$ &  & $53{,}331$ &  & $37{,}149$ &  & $31{,}406$ &  & $17{,}385$ &  \\
TIGER & $1{,}616$ & {\tiny\color{black!55}$1/25$} & $2{,}796$ & {\tiny\color{black!55}$1/19$} & $4{,}541$ & {\tiny\color{black!55}$1/8.2$} & $4{,}551$ & {\tiny\color{black!55}$1/6.9$} & $1{,}311$ & {\tiny\color{black!55}$1/13$} \\
MaskGR & $5{,}525$ & {\tiny\color{black!55}$1/7.4$} & $5{,}547$ & {\tiny\color{black!55}$1/9.6$} & $1{,}344$ & {\tiny\color{black!55}$1/28$} & $5{,}552$ & {\tiny\color{black!55}$1/5.7$} & $5{,}995$ & {\tiny\color{black!55}$1/2.9$} \\
\bottomrule
\end{tabular}

\end{table}

\FloatBarrier
\section{Balanced RQ-GMM}
\label{app:rqgmm}
\textbf{Why the balanced variant.} RQ-GMM~\citep{rqgmm2026} fits each residual level by soft EM~\citep{DEMP1977},
which places no constraint on how item mass is distributed over the $K$ components, and in some of our experiments a
few components absorb most of the catalog. Because an item's identifier is the $L$-tuple $(c^{(1)}_i,\dots,c^{(L)}_i)$, an
unbalanced level shrinks the effective identifier space, so more items collide on the same tuple and the clustering
the SID is meant to provide degrades.
We therefore replace the soft EM by a hard-assignment classification EM, which imposes the constraint
where the decision is actually made. The soft responsibility then plays no part in the fit at all: it survives only
as a read-time quantity, the prior in \eqnref{eq:gmmresp}.

\textbf{The quantizer.} Let $\{\mathbf{q}_i\}_{i=1}^{N}$ be the item embeddings and
$\mathbf{e}^{(1)}_i=\mathbf{q}_i$. Level $s$ models the residual left by the levels above it with a $K$-component
diagonal Gaussian mixture whose means form that level's codebook, and passes on
$\mathbf{e}^{(s+1)}_i=\mathbf{e}^{(s)}_i-\boldsymbol\mu^{(s)}_{c^{(s)}_i}$. Writing $\omega^{(s)}_k$ for the mixing
coefficients, the cost of placing a residual $\mathbf{e}$ in component $k$ is its negative log-likelihood,
\begin{equation}
\psi^{(s)}_k(\mathbf{e})=-\log\omega^{(s)}_k
  +\tfrac12\sum_{j=1}^{d}\Big[\frac{(e_j-\mu^{(s)}_{k,j})^2}{\sigma^{(s)2}_{k,j}}+\log\sigma^{(s)2}_{k,j}\Big],
\label{eq:gmmnll}
\end{equation}
up to a constant that does not depend on $k$. Each level is fitted by a \emph{classification}
EM~\citep{celeux1992classification}, which alternates between a hard assignment and a maximum-likelihood update, and it is
the assignment step that carries the size constraint:
\begin{equation}
\min_{z\in\{0,1\}^{N\times K}}\ \sum_{i=1}^{N}\sum_{k=1}^{K} z_{ik}\,\psi^{(s)}_k(\mathbf{e}^{(s)}_i)
\quad\text{s.t.}\quad \sum_{k} z_{ik}=1\ \ \forall i,\qquad
n_{\min}\le\sum_{i} z_{ik}\le n_{\max}\ \ \forall k,
\label{eq:balassign}
\end{equation}
with $n_{\min}=\max(1,\lfloor N/K\rfloor-1)$ and $n_{\max}=\lfloor N/K\rfloor+1$. Without the size bounds,
\eqnref{eq:balassign} separates per item and returns the maximum a posteriori (MAP) code; with them it is a transportation problem, which we
solve exactly as a minimum-cost flow. This extends the constrained $k$-means of
\citet{bradley2000constrained}, who impose a minimum cluster size and solve the resulting assignment step as a
minimum-cost flow. We add the upper bound $n_{\max}$ and use the mixture cost \eqnref{eq:gmmnll} in place of squared
Euclidean distance, so the assignment remains an exact minimum-cost flow while the cost stays a likelihood.

The M-step is the maximum-likelihood update of a diagonal mixture under the hard partition
$\mathcal{C}_k=\{i: z_{ik}=1\}$,
\begin{equation}
\boldsymbol\mu^{(s)}_k=\frac{1}{|\mathcal{C}_k|}\sum_{i\in\mathcal{C}_k}\mathbf{e}^{(s)}_i,\quad
\boldsymbol\sigma^{(s)2}_k=\frac{1}{|\mathcal{C}_k|}\sum_{i\in\mathcal{C}_k}\big(\mathbf{e}^{(s)}_i-\boldsymbol\mu^{(s)}_k\big)^{2}+\varepsilon,\quad
\omega^{(s)}_k=\frac{|\mathcal{C}_k|}{N},
\label{eq:mstep}
\end{equation}
with a covariance ridge $\varepsilon=10^{-6}$ added to every variance so \eqnref{eq:gmmnll} does not send the cost to infinity on coordinates without spread. Each level is initialized with one short run of unconstrained $k$-means
($k$-means\texttt{++} seeding~\citep{arthur2007kmeanspp}, at most $50$ Lloyd steps~\citep{lloyd1982least}) and
iterated until the assigned cost
$\sum_i\psi^{(s)}_{c^{(s)}_i}(\mathbf{e}^{(s)}_i)$ changes by less than $10^{-4}$ relatively or fewer than $0.1\%$ of
items are reassigned, with a cap of $200$ iterations. After reaching either stopping criterion or the iteration limit, we
fix the resulting assignments, update the residuals, and fit the next level. Balance therefore holds by construction
at every level: no code group differs from $\lfloor N/K\rfloor$ by more than one, whatever the scale of the embedding.
A final de-duplication digit, ranking the items inside each colliding $L$-tuple, makes the full identifier unique.

\textbf{Use as a prior.} The hard RQ-SID baseline consumes the codes $c^{(s)}_i$ directly. \methodname\ instead
consumes the level-$s$ responsibility, the posterior over the $K$ components given the residual,
\begin{equation}
\gamma^{(s)}_{i}=\mathrm{softmax}_k\!\big(-\psi^{(s)}_k(\mathbf{e}^{(s)}_i)\big),
\qquad
\rho^{(s)}_i=\mathrm{softmax}_k\!\big(-\psi^{(s)}_k(\mathbf{e}^{(s)}_i)/\tau_\rho\big),
\label{eq:gmmresp}
\end{equation}
where $\rho^{(s)}_i$ is the tempered version that \eqnref{eq:sidprior} uses.
Since $\psi$ sums over the $d$ embedding dimensions, its spread
across components grows with $d$, so $\tau_\rho$ also scales with $d$. We use
$\tau_\rho{=}400$ at $d{=}2560$.
The residual path in \eqnref{eq:gmmresp} is the \emph{greedy} one, $\mathbf{e}^{(s+1)}_i=\mathbf{e}^{(s)}_i-\boldsymbol\mu^{(s)}_{k^\ast}$ with $k^\ast=\arg\max_k\gamma^{(s)}_i[k]$.
Thus, the mode of $\gamma^{(s)}_i$ is not always the stored code $c^{(s)}_i$, although they agree for most items. For example, on Beauty, they coincide for $92$--$94\%$ of items at each level.
Feeding the responsibility rather than the code is what lets the tokenizer prior carry its uncertainty.

\FloatBarrier
\section{The two priors}
\label{app:priors}

\methodname\ injects external knowledge in two places: a \emph{tokenizer} prior on which codes are assigned to an item
(\eqnref{eq:sidprior}) and a \emph{co-visitation} prior on which item to retrieve (\eqnref{eq:decode}). Both follow the
same rule, which this appendix states once, justifies, and then instantiates in both settings.

\textbf{The rule.} Let $q_\theta\propto\exp(\ell)$ be a softmax over a finite set with learned logits $\ell$, and
let $\pi$ be a prior belief over that same set. Multiplying the two and renormalizing,
\begin{equation}
p\ \propto\ \underbrace{\exp(\ell)}_{\text{evidence}}\cdot\ \underbrace{\pi^{\,w}}_{\text{prior}},
\qquad\text{equivalently}\qquad
\log p=\ell+w\log\pi+\mathrm{const},
\label{eq:pool}
\end{equation}
so \emph{adding a scaled log-prior to the logits is equivalent to multiplying the two beliefs}, with $w$ weighting the prior.
This is log-linear pooling. The trust $w\ge0$ interpolates from
discarding the prior ($w{=}0$) to collapsing $p$ onto $\arg\max\pi$ ($w\!\to\!\infty$).

\textbf{Why not mix probabilities.}
What separates \eqnref{eq:pool} from the obvious alternative $p=(1-\beta)q_\theta+\beta\pi$ is that any finite $w$ leaves the hypothesis class unchanged. That is, provided the prior has full support, \eqnref{eq:sidprior} can be overridden by evidence:
the learned logits can add an arbitrary vector, so every interior point of $\Delta$ stays reachable. The persistent prior therefore biases the parameterization and optimization without restricting the reachable distributions.
To ensure full support, we add $\epsilon_\rho$ inside the log of \eqnref{eq:sidprior} and mix the graph prior with the uniform distribution in \eqnref{eq:graphprior}. A prior that is exactly zero on an outcome removes it at any $w>0$.
In contrast, the alternative linear mixing at a fixed $\beta>0$ restricts $p$, since $p_k\ge\beta\pi_k$ no matter the learned evidence.

\textbf{Instance 1: the tokenizer prior on codes.}
\label{app:sidprior}
At depth $t$, the learned evidence for chain $s$ is $\ell=\phi_{s,t}(\f_i)$. The tokenizer supplies the same prior
to every chain at that depth: the responsibility that the fitted quantizer places on each of its $K$ Gaussians for
item $i$, denoted $\gamma^{(t)}_{i,k}$ (Appendix~\ref{app:rqgmm}). This is not a hand-built pseudo-count but an exact
posterior under a fitted GMM generative model. It is read along a greedy residual path rather than the balanced
assignment the quantizer was fitted with, so its mode need not be the stored code.

We use the tempered responsibility $\rho_i^{(t)}$ defined in \eqnref{eq:gmmresp} as the prior in
\eqnref{eq:sidprior}, with $\alpha_t$ controlling its influence.

\textbf{Instance 2: the co-visitation prior on items.}
\label{app:graphprior}
Here the set is the catalog $\I$, the evidence is the inner product $\ell=\mathbf{u}^\top\x_i$ against the shared item
table, and the prior is $\pi=\tilde g(i\mid H)$ with trust $w$, yielding \eqnref{eq:decode}. \eqnref{eq:decode}
thus makes explicit a prior that a plain dot-product softmax leaves implicit.
We build $\tilde g$
once from the training sequences in three steps:
\begin{equation}
P=D^{-1}C,\quad D=\mathrm{diag}(C\mathbf{1});
\qquad
\hat g_H=\frac{q_H\,P^{K_s}}{\|q_H\,P^{K_s}\|_1};
\qquad
\tilde g=(1-\beta)\,\tfrac{1}{|\I|}\mathbf{1}+\beta\,\hat g_H .
\label{eq:graphprior}
\end{equation}
First, $C_{ij}$ counts how often items $i$ and $j$ occur in the same training sequence, and row-normalizing turns
those counts into conditional frequencies: $P_{ij}$ estimates how likely $j$ is to appear given that $i$ does, so
every row is a distribution over the catalog and $P$ is the transition kernel of a random walk on the item graph.
Second, we summarize the history by a \emph{recency-weighted} distribution $q_H$, which puts mass only on items with which the
user interacted, weighted according to recency: the item at age $a$ (most recent $a{=}0$) gets weight
$\propto\eta^{\,a}$, repeats accumulate, and the result is $L_1$-normalized. Applying the kernel $K_s$ times lets that
mass flow outward along co-occurrence edges as a $K_s$-step random walk, so $\hat g_H[j]$ is large
when $j$ co-occurs strongly with what the user recently touched. Third, a raw $\hat g_H$ can place zero mass on an
unseen (context, item) pair, which would veto that item outright, however strong the evidence. Thus, to ensure full support, we temper it
toward the uniform distribution. For $0\le\beta<1$, this keeps $\log\tilde g$ finite; $\beta{=}0$ ignores the graph,
whereas $\beta{=}1$ trusts it fully but can contain zero-probability outcomes.

Only $w$ and $\beta$ are tuned; all other settings use fixed defaults that are identical everywhere and never searched: whole-sequence symmetric counts with no self-loops, $\eta{=}0.5$, $K_s{=}1$ (so only direct neighbors
receive mass), and the same truncated history the encoder reads, which at $m{=}3$ gives $q_H=(0.571,\,0.286,\,0.143)$ from newest to oldest.
The same log-prior is added to the training logits of both
cross-entropies, so the model learns to complement the prior.

\FloatBarrier
\section{Gradient coupling induced by prototype sharing}
\label{app:gradient}

Let $V=|\I|$, and stack the item embeddings and item-specific vectors into $\mX,R\in\R^{V\times D}$. For chain
$s$ and depth $t$, let $A^{(s,t)}\in\R^{V\times K}$ and $C^{(s,t)}\in\R^{K\times D}$. To isolate the direct
effect of prototype sharing, hold the assignments fixed and consider a loss component $\mathcal L(\mX)$ whose
dependence on $R$ and $C^{(s,t)}$ is only through $\mX$. Let $G=\nabla_{\mX}\mathcal L$.
\begin{equation}
\mX=R+\frac{\gamma}{S}\sum_{s,t}A^{(s,t)}C^{(s,t)},\qquad
\nabla_{C^{(s,t)}}\mathcal L=\frac{\gamma}{S}A^{(s,t)\top}G,\qquad
\nabla_R\mathcal L=G.
\label{eq:appendix-gradients}
\end{equation}
Consider one gradient-descent step with learning rate $\eta$, and let $\Delta\mX$ denote the resulting change in the
item table. \eqnref{eq:appendix-gradients} gives
\begin{equation}
\Delta R=-\eta G,
\qquad
\Delta C^{(s,t)}=-\eta\frac{\gamma}{S}A^{(s,t)\top}G.
\label{eq:parameter-gradient-flow}
\end{equation}
Because the assignments are fixed, substituting these parameter updates into the composed table gives
\begin{equation}
\begin{aligned}
\Delta\mX
&=\Delta R+\frac{\gamma}{S}\sum_{s,t}A^{(s,t)}\Delta C^{(s,t)}\\
&=-\eta G-\eta\frac{\gamma^2}{S^2}\sum_{s,t}A^{(s,t)}A^{(s,t)\top}G\\
&=-\eta(I_V+\mathcal K)G,
\qquad
\mathcal K=\frac{\gamma^2}{S^2}\sum_{s,t}A^{(s,t)}A^{(s,t)\top},
\end{aligned}
\label{eq:gradient-kernel}
\end{equation}
where $I_V\in\R^{V\times V}$ is the identity matrix. Consequently,
\begin{equation}
\mathcal K_{ij}=\frac{\gamma^2}{S^2}\sum_{s,t}
\left\langle A_i^{(s,t)},A_j^{(s,t)}\right\rangle,
\label{eq:gradient-overlap}
\end{equation}
which couples item updates in proportion to assignment overlap, while the identity term $I_V$ preserves a direct
item-specific update. This analysis isolates the direct representation path with fixed assignments, while the complete
model additionally learns the assignments and contains losses that depend explicitly on the prototypes.

\FloatBarrier
\section{Diagnostic results}
\label{app:diagnostics}

Table~\ref{tab:motivation} separates the semantic information in the RQ-SID assignments from the parameter sharing
induced by the codebook. The random control preserves the table structure, depth, codebook size, and codeword
vectors while randomly permuting the item-to-code assignments.

\begin{table}[tbp]\centering\small
\caption{Item-table variants for a continuous flow recommender. The flat table of
each block is absolute test HR@5~$\uparrow$; every other cell is its \% change relative to it. A hard SID beats random codes
from scratch, but with the full content embedding the ordering reverses and, on the sparse catalogs, both fall below the
trainable flat table.}
\label{tab:motivation}
\providecommand{\pending}[1]{\textcolor{red}{#1}}
\providecommand{\scen}[1]{\textsc{#1}}
\providecommand{\armname}[1]{\textsc{#1}}

\begin{tabular}{@{}ll ccc@{}}
\toprule
scenario & item table & Beauty & Industrial & Steam \\
\midrule
\multirow{3}{*}{\scen{from scratch}} & \armname{flat}   & $5.77$ & $2.75$ & $8.29$ \\
                              & \;$+$ \armname{hard} RQ-SID   & $+0.52\%$ & $+9.82\%$ & $+0.36\%$ \\
                              & \;$+$ \armname{random} SID    & $-1.73\%$ & $+0.73\%$ & $+0.12\%$ \\
\midrule
\multirow{4}{*}{\scen{PCA-text}} & \armname{flat}, trainable & $6.69$ & $3.81$ & $8.13$ \\
                                & \;\armname{frozen}             & $-18.24\%$ & $-20.21\%$ & $-14.15\%$ \\
                                & \;$+$ \armname{hard} RQ-SID  & $-3.89\%$ & $-7.61\%$ & $+0.74\%$ \\
                                & \;$+$ \armname{random} SID   & $+0.30\%$ & $-1.84\%$ & $+0.86\%$ \\
\bottomrule
\end{tabular}

\end{table}

In the \scen{from scratch} regime, the hard RQ-SID table improves HR@5~$\uparrow$ over the flat table on all three datasets, whereas
matched random assignments yield smaller gains or a loss. This isolates useful semantic structure in the
pretrained assignments beyond parameter sharing alone. With \scen{PCA-text} initialization, freezing the content
embeddings causes losses ranging from \FrozenRangeLo\ to \FrozenRangeHi. On the two Amazon datasets, hard semantic assignments also
underperform both the trainable flat table and the random-assignment control, suggesting that a fixed semantic
partition can conflict with the recommendation objective once the continuous content representation is available.
This motivates using the tokenizer to initialize differentiable assignments that can be optimized jointly with
the recommender. On Steam, PCA initialization itself reduces performance, while hard and random assignments
produce similar gains below $1\%$; we hypothesize that the content embedding is less informative for next-item
prediction on this dataset, causing its semantic partition to behave similarly to random sharing.

\textbf{Removing the co-visitation prior.}
Table~\ref{tab:noprior} repeats the scenario study with the co-visitation prior disabled for every variant. On Beauty
and Industrial, \methodname\ remains above the corresponding baseline in all three regimes. On Steam, it is
essentially flat relative to the baseline in \scen{from scratch} and \scen{PCA-text}, but retains a \NoPriorSidSteam\
HR@5~$\uparrow$ advantage over the hard table in \scen{offline SID}.
These results show that the gradient-pooling benefit from our prototype mixtures is complementary to the retrieval prior.

\begin{table}[tbp]
\centering\small
\caption{\textbf{Controlled results without the co-visitation prior.} All values are test HR@5~$\uparrow$
($\times100$). Each variant's cell gives its absolute score over its relative change from the from-scratch flat table.
Each \emph{vs.}\ column compares \methodname\ with that regime's baseline.}
\label{tab:noprior}
\resizebox{\linewidth}{!}{\providecommand{\pending}[1]{\textcolor{red}{#1}}
\providecommand{\scen}[1]{\textsc{#1}}
\providecommand{\armname}[1]{\textsc{#1}}

\begin{tabular}{@{}l ccc ccc cccc@{}}
\toprule
 & \multicolumn{3}{c}{\textbf{\scen{from scratch}}} & \multicolumn{3}{c}{\textbf{\scen{offline SID}}} & \multicolumn{4}{c}{\textbf{\scen{PCA-text}}} \\
\cmidrule(lr){2-4}\cmidrule(lr){5-7}\cmidrule(lr){8-11}
dataset & \armname{flat} & \methodname & \emph{vs.} & \armname{hard} & \methodname & \emph{vs.} & \armname{flat} & \armname{hard} & \methodname & \emph{vs.} \\
\midrule
Beauty & \makecell{$5.61$\\{\scriptsize\emph{ref.}}} & \makecell{$5.67$\\{\scriptsize$+1.07\%$}} & $+1.07\%$ & \makecell{$5.76$\\{\scriptsize$+2.67\%$}} & \makecell{$5.89$\\{\scriptsize$+4.99\%$}} & $+2.26\%$ & \makecell{$6.27$\\{\scriptsize$+11.76\%$}} & \makecell{$6.27$\\{\scriptsize$+11.76\%$}} & \makecell{$6.50$\\{\scriptsize$+15.86\%$}} & $+3.67\%$ \\
Industrial & \makecell{$2.64$\\{\scriptsize\emph{ref.}}} & \makecell{$2.81$\\{\scriptsize$+6.44\%$}} & $+6.44\%$ & \makecell{$2.85$\\{\scriptsize$+7.95\%$}} & \makecell{$3.21$\\{\scriptsize$+21.59\%$}} & $+12.63\%$ & \makecell{$3.62$\\{\scriptsize$+37.12\%$}} & \makecell{$3.51$\\{\scriptsize$+32.95\%$}} & \makecell{$3.68$\\{\scriptsize$+39.39\%$}} & $+1.66\%$ \\
Steam & \makecell{$8.28$\\{\scriptsize\emph{ref.}}} & \makecell{$8.26$\\{\scriptsize$-0.24\%$}} & $-0.24\%$ & \makecell{$7.96$\\{\scriptsize$-3.86\%$}} & \makecell{$8.51$\\{\scriptsize$+2.78\%$}} & $+6.91\%$ & \makecell{$8.10$\\{\scriptsize$-2.17\%$}} & \makecell{$8.21$\\{\scriptsize$-0.85\%$}} & \makecell{$8.14$\\{\scriptsize$-1.69\%$}} & $+0.49\%$ \\
\bottomrule
\end{tabular}
}
\end{table}

\textbf{Frequency stratification.}
We group test users by the training frequency of their target item, without splitting users whose targets have the
same frequency. The resulting approximately equal-mass groups contain $22{,}363$ Beauty users (rare: at most $6$
interactions; medium: $7$--$22$; frequent: more than $22$) and $51{,}322$ Industrial users (rare: at most $7$;
medium: $8$--$28$; frequent: more than $28$). Table~\ref{tab:frequency-summary} reports the relative gain within
each group.

\begin{table}[tbp]
\centering\small
\caption{\textbf{Relative gains by target-item training frequency.} Test users are grouped by their target item's
frequency in the training set. Entries are the relative improvement of \methodname\ over the indicated baseline.
Relative gains use unrounded scores and may differ slightly from calculations using displayed values.}
\label{tab:frequency-summary}
\setlength{\tabcolsep}{4pt}
\resizebox{\linewidth}{!}{%
\begin{tabular}{@{}lllrrrrl@{}}
\toprule
scenario & baseline & dataset and metric & rare & medium & frequent & all & larger end \\
\midrule
\scen{from scratch} & \armname{flat} & Beauty HR@5 $\uparrow$ & $+0.2\%$ & $-1.8\%$ & $+4.1\%$ & $+2.1\%$ & frequent \\
 &  & Beauty NDCG@10 $\uparrow$ & $-3.3\%$ & $-1.5\%$ & $+2.9\%$ & $+0.9\%$ & frequent \\
 &  & Industrial HR@5 $\uparrow$ & $+5.9\%$ & $+4.5\%$ & $+4.4\%$ & $+4.5\%$ & rare \\
 &  & Industrial NDCG@10 $\uparrow$ & $+4.6\%$ & $+1.8\%$ & $+5.1\%$ & $+4.5\%$ & frequent \\
\midrule
\scen{offline SID} & \armname{hard} & Beauty HR@5 $\uparrow$ & $+8.5\%$ & $+8.8\%$ & $+8.9\%$ & $+8.8\%$ & frequent \\
 &  & Beauty NDCG@10 $\uparrow$ & $+10.9\%$ & $+6.6\%$ & $+4.8\%$ & $+6.1\%$ & rare \\
 &  & Industrial HR@5 $\uparrow$ & $+49.7\%$ & $+26.6\%$ & $+10.9\%$ & $+14.7\%$ & rare \\
 &  & Industrial NDCG@10 $\uparrow$ & $+56.0\%$ & $+32.3\%$ & $+17.1\%$ & $+20.8\%$ & rare \\
\midrule
\scen{PCA-text} & \armname{flat} & Beauty HR@5 $\uparrow$ & $+5.2\%$ & $+3.6\%$ & $-0.3\%$ & $+1.4\%$ & rare \\
 &  & Beauty NDCG@10 $\uparrow$ & $+4.2\%$ & $+3.1\%$ & $-0.4\%$ & $+1.1\%$ & rare \\
 &  & Industrial HR@5 $\uparrow$ & $+7.1\%$ & $+2.6\%$ & $-1.4\%$ & $-0.3\%$ & rare \\
 &  & Industrial NDCG@10 $\uparrow$ & $+9.0\%$ & $+0.8\%$ & $-2.5\%$ & $-1.4\%$ & rare \\
\midrule
\scen{PCA-text} & \armname{hard} & Beauty HR@5 $\uparrow$ & $+10.7\%$ & $+3.0\%$ & $+5.4\%$ & $+5.5\%$ & rare \\
 &  & Beauty NDCG@10 $\uparrow$ & $+5.3\%$ & $+3.2\%$ & $+5.1\%$ & $+4.6\%$ & rare \\
 &  & Industrial HR@5 $\uparrow$ & $+15.3\%$ & $+8.7\%$ & $+6.9\%$ & $+7.6\%$ & rare \\
 &  & Industrial NDCG@10 $\uparrow$ & $+18.8\%$ & $+9.7\%$ & $+8.3\%$ & $+9.0\%$ & rare \\
\bottomrule
\end{tabular}}
\end{table}

\section{Full results}
\label{app:fullresults}
Table~\ref{tab:summary-full} summarizes all evaluation cutoffs for the validation-selected configuration of each
model and dataset. Tables~\ref{tab:raw-beauty}--\ref{tab:raw-office} then give the complete per-dataset results
across all model families, initializations, and metrics.

\begin{table}[!htbp]
\centering\small
\caption{\textbf{Full cross-dataset results}, HR@$k$~$\uparrow$ and NDCG@$k$~$\uparrow$
(mean$\pm$std, $\times100$). Each model and dataset uses the configuration selected by the validation geometric
mean of HR@5~$\uparrow$ and NDCG@5~$\uparrow$; the same run supplies every cutoff. Superscripts denote
$^{\text{s}}$ scratch, $^{\text{i}}$ offline SID, and $^{\text{p}}$ PCA-text. \textbf{Bold}~=~best,
\underline{underline}~=~2nd per column, and $\Delta$ compares \methodname\ with the strongest baseline.}
\label{tab:summary-full}
\resizebox{\linewidth}{!}{\providecommand{\pending}[1]{\textcolor{red}{#1}}

\setlength{\tabcolsep}{3pt}
\renewcommand{\cellalign}{tc}
\begin{tabular}{@{}l cccccc cccccc@{}}
\toprule
 & \multicolumn{6}{c}{\textbf{Beauty}} & \multicolumn{6}{c}{\textbf{Indus.}} \\
\cmidrule(lr){2-7}\cmidrule(lr){8-13}
 & HR@5 & N@5 & HR@10 & N@10 & HR@20 & N@20 & HR@5 & N@5 & HR@10 & N@10 & HR@20 & N@20 \\
\midrule
FlowRec & \makecell{\underline{6.27}$^{\text{p}}$\\[-3pt]{\tiny$\pm$0.18}} & \makecell{\underline{4.03}\\[-3pt]{\tiny$\pm$0.03}} & \makecell{\underline{9.52}\\[-3pt]{\tiny$\pm$0.22}} & \makecell{\underline{5.07}\\[-3pt]{\tiny$\pm$0.05}} & \makecell{\underline{13.52}\\[-3pt]{\tiny$\pm$0.41}} & \makecell{\underline{6.08}\\[-3pt]{\tiny$\pm$0.10}} & \makecell{\underline{3.62}$^{\text{p}}$\\[-3pt]{\tiny$\pm$0.02}} & \makecell{2.04\\[-3pt]{\tiny$\pm$0.03}} & \makecell{\underline{5.54}\\[-3pt]{\tiny$\pm$0.04}} & \makecell{2.66\\[-3pt]{\tiny$\pm$0.04}} & \makecell{\underline{8.03}\\[-3pt]{\tiny$\pm$0.04}} & \makecell{3.28\\[-3pt]{\tiny$\pm$0.04}} \\
TIGER & \makecell{3.51$^{\text{s}}$\\[-3pt]{\tiny$\pm$0.07}} & \makecell{2.54\\[-3pt]{\tiny$\pm$0.05}} & \makecell{4.92\\[-3pt]{\tiny$\pm$0.09}} & \makecell{2.99\\[-3pt]{\tiny$\pm$0.05}} & \makecell{6.90\\[-3pt]{\tiny$\pm$0.21}} & \makecell{3.49\\[-3pt]{\tiny$\pm$0.08}} & \makecell{2.05$^{\text{p}}$\\[-3pt]{\tiny$\pm$0.03}} & \makecell{1.34\\[-3pt]{\tiny$\pm$0.02}} & \makecell{3.25\\[-3pt]{\tiny$\pm$0.03}} & \makecell{1.73\\[-3pt]{\tiny$\pm$0.02}} & \makecell{4.61\\[-3pt]{\tiny$\pm$0.01}} & \makecell{2.08\\[-3pt]{\tiny$\pm$0.01}} \\
MaskGR & \makecell{5.06$^{\text{i}}$\\[-3pt]{\tiny$\pm$0.13}} & \makecell{3.28\\[-3pt]{\tiny$\pm$0.12}} & \makecell{8.11\\[-3pt]{\tiny$\pm$0.15}} & \makecell{4.27\\[-3pt]{\tiny$\pm$0.12}} & \makecell{12.10\\[-3pt]{\tiny$\pm$0.16}} & \makecell{5.29\\[-3pt]{\tiny$\pm$0.09}} & \makecell{3.46$^{\text{p}}$\\[-3pt]{\tiny$\pm$0.03}} & \makecell{\underline{2.24}\\[-3pt]{\tiny$\pm$0.03}} & \makecell{5.25\\[-3pt]{\tiny$\pm$0.06}} & \makecell{\underline{2.82}\\[-3pt]{\tiny$\pm$0.02}} & \makecell{7.46\\[-3pt]{\tiny$\pm$0.07}} & \makecell{\underline{3.38}\\[-3pt]{\tiny$\pm$0.01}} \\
\textbf{\methodname} & \makecell{\textbf{6.79}$^{\text{p}}$\\[-3pt]{\tiny$\pm$0.06}} & \makecell{\textbf{4.73}\\[-3pt]{\tiny$\pm$0.04}} & \makecell{\textbf{9.75}\\[-3pt]{\tiny$\pm$0.06}} & \makecell{\textbf{5.69}\\[-3pt]{\tiny$\pm$0.04}} & \makecell{\textbf{13.53}\\[-3pt]{\tiny$\pm$0.17}} & \makecell{\textbf{6.64}\\[-3pt]{\tiny$\pm$0.07}} & \makecell{\textbf{3.79}$^{\text{p}}$\\[-3pt]{\tiny$\pm$0.01}} & \makecell{\textbf{2.34}\\[-3pt]{\tiny$\pm$0.03}} & \makecell{\textbf{5.77}\\[-3pt]{\tiny$\pm$0.02}} & \makecell{\textbf{2.98}\\[-3pt]{\tiny$\pm$0.02}} & \makecell{\textbf{8.36}\\[-3pt]{\tiny$\pm$0.08}} & \makecell{\textbf{3.63}\\[-3pt]{\tiny$\pm$0.04}} \\
$\Delta$ (\%) & \textcolor{green!55!black}{$+8.3\%$} & \textcolor{green!55!black}{$+17.4\%$} & \textcolor{green!55!black}{$+2.4\%$} & \textcolor{green!55!black}{$+12.2\%$} & \textcolor{green!55!black}{$+0.1\%$} & \textcolor{green!55!black}{$+9.2\%$} & \textcolor{green!55!black}{$+4.7\%$} & \textcolor{green!55!black}{$+4.5\%$} & \textcolor{green!55!black}{$+4.2\%$} & \textcolor{green!55!black}{$+5.7\%$} & \textcolor{green!55!black}{$+4.1\%$} & \textcolor{green!55!black}{$+7.4\%$} \\
\midrule
 & \multicolumn{6}{c}{\textbf{Steam}} & \multicolumn{6}{c}{\textbf{ML-1M}} \\
\cmidrule(lr){2-7}\cmidrule(lr){8-13}
 & HR@5 & N@5 & HR@10 & N@10 & HR@20 & N@20 & HR@5 & N@5 & HR@10 & N@10 & HR@20 & N@20 \\
\midrule
FlowRec & \makecell{8.28$^{\text{s}}$\\[-3pt]{\tiny$\pm$0.33}} & \makecell{\underline{5.37}\\[-3pt]{\tiny$\pm$0.30}} & \makecell{13.54\\[-3pt]{\tiny$\pm$0.31}} & \makecell{7.06\\[-3pt]{\tiny$\pm$0.29}} & \makecell{20.61\\[-3pt]{\tiny$\pm$0.35}} & \makecell{8.84\\[-3pt]{\tiny$\pm$0.30}} & \makecell{15.36$^{\text{p}}$\\[-3pt]{\tiny$\pm$0.20}} & \makecell{10.14\\[-3pt]{\tiny$\pm$0.12}} & \makecell{22.94\\[-3pt]{\tiny$\pm$0.04}} & \makecell{12.58\\[-3pt]{\tiny$\pm$0.05}} & \makecell{32.68\\[-3pt]{\tiny$\pm$0.14}} & \makecell{15.03\\[-3pt]{\tiny$\pm$0.09}} \\
TIGER & \makecell{8.43$^{\text{i}}$\\[-3pt]{\tiny$\pm$0.10}} & \makecell{5.01\\[-3pt]{\tiny$\pm$0.03}} & \makecell{13.28\\[-3pt]{\tiny$\pm$0.63}} & \makecell{6.57\\[-3pt]{\tiny$\pm$0.19}} & \makecell{20.46\\[-3pt]{\tiny$\pm$1.07}} & \makecell{8.36\\[-3pt]{\tiny$\pm$0.24}} & \makecell{12.59$^{\text{s}}$\\[-3pt]{\tiny$\pm$0.40}} & \makecell{8.47\\[-3pt]{\tiny$\pm$0.28}} & \makecell{19.03\\[-3pt]{\tiny$\pm$0.37}} & \makecell{10.54\\[-3pt]{\tiny$\pm$0.28}} & \makecell{27.20\\[-3pt]{\tiny$\pm$0.68}} & \makecell{12.61\\[-3pt]{\tiny$\pm$0.36}} \\
MaskGR & \makecell{\underline{8.64}$^{\text{p}}$\\[-3pt]{\tiny$\pm$0.00}} & \makecell{\textbf{5.57}\\[-3pt]{\tiny$\pm$0.01}} & \makecell{\textbf{14.43}\\[-3pt]{\tiny$\pm$0.05}} & \makecell{\textbf{7.42}\\[-3pt]{\tiny$\pm$0.02}} & \makecell{\textbf{20.97}\\[-3pt]{\tiny$\pm$0.23}} & \makecell{\textbf{9.08}\\[-3pt]{\tiny$\pm$0.07}} & \makecell{\underline{15.64}$^{\text{s}}$\\[-3pt]{\tiny$\pm$0.40}} & \makecell{\underline{10.23}\\[-3pt]{\tiny$\pm$0.28}} & \makecell{\underline{23.61}\\[-3pt]{\tiny$\pm$0.29}} & \makecell{\underline{12.79}\\[-3pt]{\tiny$\pm$0.26}} & \makecell{\underline{33.93}\\[-3pt]{\tiny$\pm$0.17}} & \makecell{\underline{15.39}\\[-3pt]{\tiny$\pm$0.21}} \\
\textbf{\methodname} & \makecell{\textbf{8.69}$^{\text{i}}$\\[-3pt]{\tiny$\pm$0.29}} & \makecell{\textbf{5.57}\\[-3pt]{\tiny$\pm$0.19}} & \makecell{\underline{13.94}\\[-3pt]{\tiny$\pm$0.32}} & \makecell{\underline{7.25}\\[-3pt]{\tiny$\pm$0.20}} & \makecell{\underline{20.82}\\[-3pt]{\tiny$\pm$0.42}} & \makecell{\underline{8.99}\\[-3pt]{\tiny$\pm$0.23}} & \makecell{\textbf{17.25}$^{\text{p}}$\\[-3pt]{\tiny$\pm$0.13}} & \makecell{\textbf{11.71}\\[-3pt]{\tiny$\pm$0.11}} & \makecell{\textbf{24.71}\\[-3pt]{\tiny$\pm$0.38}} & \makecell{\textbf{14.11}\\[-3pt]{\tiny$\pm$0.17}} & \makecell{\textbf{33.98}\\[-3pt]{\tiny$\pm$0.34}} & \makecell{\textbf{16.44}\\[-3pt]{\tiny$\pm$0.15}} \\
$\Delta$ (\%) & \textcolor{green!55!black}{$+0.6\%$} & $0.0\%$ & \textcolor{red!70!black}{$-3.4\%$} & \textcolor{red!70!black}{$-2.3\%$} & \textcolor{red!70!black}{$-0.7\%$} & \textcolor{red!70!black}{$-1.0\%$} & \textcolor{green!55!black}{$+10.3\%$} & \textcolor{green!55!black}{$+14.5\%$} & \textcolor{green!55!black}{$+4.7\%$} & \textcolor{green!55!black}{$+10.3\%$} & \textcolor{green!55!black}{$+0.1\%$} & \textcolor{green!55!black}{$+6.8\%$} \\
\midrule
 & \multicolumn{6}{c}{\textbf{Yelp}} & \multicolumn{6}{c}{\textbf{Office}} \\
\cmidrule(lr){2-7}\cmidrule(lr){8-13}
 & HR@5 & N@5 & HR@10 & N@10 & HR@20 & N@20 & HR@5 & N@5 & HR@10 & N@10 & HR@20 & N@20 \\
\midrule
FlowRec & \makecell{2.98$^{\text{p}}$\\[-3pt]{\tiny$\pm$0.05}} & \makecell{1.86\\[-3pt]{\tiny$\pm$0.03}} & \makecell{\underline{5.10}\\[-3pt]{\tiny$\pm$0.05}} & \makecell{\underline{2.54}\\[-3pt]{\tiny$\pm$0.03}} & \makecell{\underline{8.34}\\[-3pt]{\tiny$\pm$0.06}} & \makecell{\underline{3.35}\\[-3pt]{\tiny$\pm$0.04}} & \makecell{\underline{2.90}$^{\text{p}}$\\[-3pt]{\tiny$\pm$0.02}} & \makecell{\underline{1.67}\\[-3pt]{\tiny$\pm$0.02}} & \makecell{\textbf{4.41}\\[-3pt]{\tiny$\pm$0.04}} & \makecell{\underline{2.15}\\[-3pt]{\tiny$\pm$0.02}} & \makecell{\textbf{6.23}\\[-3pt]{\tiny$\pm$0.02}} & \makecell{\underline{2.61}\\[-3pt]{\tiny$\pm$0.02}} \\
TIGER & \makecell{1.29$^{\text{p}}$\\[-3pt]{\tiny$\pm$0.07}} & \makecell{0.80\\[-3pt]{\tiny$\pm$0.04}} & \makecell{2.09\\[-3pt]{\tiny$\pm$0.09}} & \makecell{1.06\\[-3pt]{\tiny$\pm$0.04}} & \makecell{3.03\\[-3pt]{\tiny$\pm$0.10}} & \makecell{1.30\\[-3pt]{\tiny$\pm$0.04}} & \makecell{1.52$^{\text{p}}$\\[-3pt]{\tiny$\pm$0.01}} & \makecell{0.98\\[-3pt]{\tiny$\pm$0.01}} & \makecell{2.33\\[-3pt]{\tiny$\pm$0.03}} & \makecell{1.25\\[-3pt]{\tiny$\pm$0.00}} & \makecell{3.34\\[-3pt]{\tiny$\pm$0.08}} & \makecell{1.50\\[-3pt]{\tiny$\pm$0.02}} \\
MaskGR & \makecell{\underline{2.99}$^{\text{s}}$\\[-3pt]{\tiny$\pm$0.04}} & \makecell{\underline{1.87}\\[-3pt]{\tiny$\pm$0.03}} & \makecell{\underline{5.10}\\[-3pt]{\tiny$\pm$0.07}} & \makecell{\underline{2.54}\\[-3pt]{\tiny$\pm$0.03}} & \makecell{8.24\\[-3pt]{\tiny$\pm$0.10}} & \makecell{3.33\\[-3pt]{\tiny$\pm$0.04}} & \makecell{2.29$^{\text{s}}$\\[-3pt]{\tiny$\pm$0.02}} & \makecell{1.53\\[-3pt]{\tiny$\pm$0.01}} & \makecell{\underline{3.40}\\[-3pt]{\tiny$\pm$0.03}} & \makecell{1.89\\[-3pt]{\tiny$\pm$0.02}} & \makecell{4.88\\[-3pt]{\tiny$\pm$0.03}} & \makecell{2.26\\[-3pt]{\tiny$\pm$0.01}} \\
\textbf{\methodname} & \makecell{\textbf{3.18}$^{\text{p}}$\\[-3pt]{\tiny$\pm$0.02}} & \makecell{\textbf{2.03}\\[-3pt]{\tiny$\pm$0.02}} & \makecell{\textbf{5.25}\\[-3pt]{\tiny$\pm$0.04}} & \makecell{\textbf{2.70}\\[-3pt]{\tiny$\pm$0.03}} & \makecell{\textbf{8.38}\\[-3pt]{\tiny$\pm$0.05}} & \makecell{\textbf{3.48}\\[-3pt]{\tiny$\pm$0.03}} & \makecell{\textbf{3.22}$^{\text{p}}$\\[-3pt]{\tiny$\pm$0.01}} & \makecell{\textbf{2.28}\\[-3pt]{\tiny$\pm$0.01}} & \makecell{\textbf{4.41}\\[-3pt]{\tiny$\pm$0.01}} & \makecell{\textbf{2.66}\\[-3pt]{\tiny$\pm$0.01}} & \makecell{\underline{5.84}\\[-3pt]{\tiny$\pm$0.03}} & \makecell{\textbf{3.02}\\[-3pt]{\tiny$\pm$0.01}} \\
$\Delta$ (\%) & \textcolor{green!55!black}{$+6.4\%$} & \textcolor{green!55!black}{$+8.6\%$} & \textcolor{green!55!black}{$+2.9\%$} & \textcolor{green!55!black}{$+6.3\%$} & \textcolor{green!55!black}{$+0.5\%$} & \textcolor{green!55!black}{$+3.9\%$} & \textcolor{green!55!black}{$+11.0\%$} & \textcolor{green!55!black}{$+36.5\%$} & $0.0\%$ & \textcolor{green!55!black}{$+23.7\%$} & \textcolor{red!70!black}{$-6.3\%$} & \textcolor{green!55!black}{$+15.7\%$} \\
\bottomrule
\end{tabular}
}
\end{table}

\begin{table}[tbp]
\centering
\caption{\textbf{Beauty --- full results.} All families $\times$ initializations. \textbf{Bold} = best,
\underline{underline} = 2nd per column within an initialization block; \methodname{} shows $\Delta\%$ vs.\ the
best baseline. FlowRec has no \scen{offline SID} variant, since a flat table has no codes to consume, so the row
marked $^{\dagger}$ repeats its \scen{from scratch} numbers.}
\label{tab:raw-beauty}
\resizebox{\linewidth}{!}{\providecommand{\pending}[1]{\textcolor{red}{#1}}

\begin{tabular}{llcccccc}
\toprule
\textbf{beauty} & \textbf{model} & HR@5 & HR@10 & HR@20 & N@5 & N@10 & N@20 \\
\midrule
scratch & FlowRec & \underline{5.61}{\tiny$\pm$0.02} & \underline{8.01}{\tiny$\pm$0.05} & \underline{10.90}{\tiny$\pm$0.20} & \underline{3.86}{\tiny$\pm$0.02} & \underline{4.63}{\tiny$\pm$0.03} & \underline{5.36}{\tiny$\pm$0.06} \\
 & TIGER & 3.51{\tiny$\pm$0.07} & 4.92{\tiny$\pm$0.09} & 6.90{\tiny$\pm$0.21} & 2.54{\tiny$\pm$0.05} & 2.99{\tiny$\pm$0.05} & 3.49{\tiny$\pm$0.08} \\
 & MaskGR & 4.38{\tiny$\pm$0.16} & 6.48{\tiny$\pm$0.19} & 9.33{\tiny$\pm$0.09} & 2.99{\tiny$\pm$0.14} & 3.66{\tiny$\pm$0.14} & 4.38{\tiny$\pm$0.12} \\
 & \textbf{\methodname} & \textbf{5.89}{\tiny$\pm$0.04}\,{\scriptsize\textcolor{green!55!black}{$+4.99\%$}} & \textbf{8.28}{\tiny$\pm$0.05}\,{\scriptsize\textcolor{green!55!black}{$+3.37\%$}} & \textbf{11.30}{\tiny$\pm$0.01}\,{\scriptsize\textcolor{green!55!black}{$+3.67\%$}} & \textbf{4.21}{\tiny$\pm$0.01}\,{\scriptsize\textcolor{green!55!black}{$+9.07\%$}} & \textbf{4.98}{\tiny$\pm$0.01}\,{\scriptsize\textcolor{green!55!black}{$+7.56\%$}} & \textbf{5.75}{\tiny$\pm$0.01}\,{\scriptsize\textcolor{green!55!black}{$+7.28\%$}} \\
\midrule
SID & FlowRec$^{\dagger}$ & \underline{5.61}{\tiny$\pm$0.02} & 8.01{\tiny$\pm$0.05} & 10.90{\tiny$\pm$0.20} & \underline{3.86}{\tiny$\pm$0.02} & \underline{4.63}{\tiny$\pm$0.03} & \underline{5.36}{\tiny$\pm$0.06} \\
 & TIGER & 3.48{\tiny$\pm$0.14} & 5.33{\tiny$\pm$0.15} & 8.00{\tiny$\pm$0.13} & 2.30{\tiny$\pm$0.14} & 2.89{\tiny$\pm$0.14} & 3.56{\tiny$\pm$0.14} \\
 & MaskGR & 5.06{\tiny$\pm$0.13} & \underline{8.11}{\tiny$\pm$0.15} & \textbf{12.10}{\tiny$\pm$0.16} & 3.28{\tiny$\pm$0.12} & 4.27{\tiny$\pm$0.12} & 5.29{\tiny$\pm$0.09} \\
 & \textbf{\methodname} & \textbf{6.32}{\tiny$\pm$0.10}\,{\scriptsize\textcolor{green!55!black}{$+12.66\%$}} & \textbf{8.72}{\tiny$\pm$0.08}\,{\scriptsize\textcolor{green!55!black}{$+7.52\%$}} & \underline{11.96}{\tiny$\pm$0.10}\,{\scriptsize\textcolor{red!70!black}{$-1.16\%$}} & \textbf{4.51}{\tiny$\pm$0.05}\,{\scriptsize\textcolor{green!55!black}{$+16.84\%$}} & \textbf{5.28}{\tiny$\pm$0.03}\,{\scriptsize\textcolor{green!55!black}{$+14.04\%$}} & \textbf{6.10}{\tiny$\pm$0.00}\,{\scriptsize\textcolor{green!55!black}{$+13.81\%$}} \\
\midrule
PCA & FlowRec & \underline{6.27}{\tiny$\pm$0.18} & \underline{9.52}{\tiny$\pm$0.22} & \underline{13.52}{\tiny$\pm$0.41} & \underline{4.03}{\tiny$\pm$0.03} & \underline{5.07}{\tiny$\pm$0.05} & \underline{6.08}{\tiny$\pm$0.10} \\
 & TIGER & 3.39{\tiny$\pm$0.09} & 5.45{\tiny$\pm$0.19} & 8.30{\tiny$\pm$0.18} & 2.24{\tiny$\pm$0.06} & 2.90{\tiny$\pm$0.09} & 3.62{\tiny$\pm$0.08} \\
 & MaskGR & 5.05{\tiny$\pm$0.13} & 7.93{\tiny$\pm$0.05} & 11.97{\tiny$\pm$0.13} & 3.26{\tiny$\pm$0.13} & 4.19{\tiny$\pm$0.10} & 5.22{\tiny$\pm$0.06} \\
 & \textbf{\methodname} & \textbf{6.79}{\tiny$\pm$0.06}\,{\scriptsize\textcolor{green!55!black}{$+8.29\%$}} & \textbf{9.75}{\tiny$\pm$0.06}\,{\scriptsize\textcolor{green!55!black}{$+2.42\%$}} & \textbf{13.53}{\tiny$\pm$0.17}\,{\scriptsize\textcolor{green!55!black}{$+0.07\%$}} & \textbf{4.73}{\tiny$\pm$0.04}\,{\scriptsize\textcolor{green!55!black}{$+17.37\%$}} & \textbf{5.69}{\tiny$\pm$0.04}\,{\scriptsize\textcolor{green!55!black}{$+12.23\%$}} & \textbf{6.64}{\tiny$\pm$0.07}\,{\scriptsize\textcolor{green!55!black}{$+9.21\%$}} \\
\bottomrule
\end{tabular}
}
\end{table}

\begin{table}[tbp]
\centering
\caption{\textbf{Industrial \& Scientific --- full results.} Conventions as in Table~\ref{tab:raw-beauty}.}
\label{tab:raw-industrial}
\resizebox{\linewidth}{!}{\providecommand{\pending}[1]{\textcolor{red}{#1}}

\begin{tabular}{llcccccc}
\toprule
\textbf{industrial} & \textbf{model} & HR@5 & HR@10 & HR@20 & N@5 & N@10 & N@20 \\
\midrule
scratch & FlowRec & \underline{2.64}{\tiny$\pm$0.00} & \underline{4.04}{\tiny$\pm$0.03} & \textbf{5.91}{\tiny$\pm$0.09} & 1.49{\tiny$\pm$0.01} & 1.94{\tiny$\pm$0.01} & 2.41{\tiny$\pm$0.02} \\
 & TIGER & 1.61{\tiny$\pm$0.05} & 2.60{\tiny$\pm$0.10} & 4.01{\tiny$\pm$0.13} & 1.05{\tiny$\pm$0.03} & 1.37{\tiny$\pm$0.05} & 1.72{\tiny$\pm$0.06} \\
 & MaskGR & 2.36{\tiny$\pm$0.02} & 3.63{\tiny$\pm$0.02} & 5.47{\tiny$\pm$0.02} & \underline{1.58}{\tiny$\pm$0.01} & \underline{1.99}{\tiny$\pm$0.01} & \underline{2.45}{\tiny$\pm$0.01} \\
 & \textbf{\methodname} & \textbf{2.88}{\tiny$\pm$0.01}\,{\scriptsize\textcolor{green!55!black}{$+9.09\%$}} & \textbf{4.15}{\tiny$\pm$0.07}\,{\scriptsize\textcolor{green!55!black}{$+2.72\%$}} & \underline{5.82}{\tiny$\pm$0.05}\,{\scriptsize\textcolor{red!70!black}{$-1.52\%$}} & \textbf{2.00}{\tiny$\pm$0.00}\,{\scriptsize\textcolor{green!55!black}{$+26.58\%$}} & \textbf{2.41}{\tiny$\pm$0.02}\,{\scriptsize\textcolor{green!55!black}{$+21.11\%$}} & \textbf{2.83}{\tiny$\pm$0.02}\,{\scriptsize\textcolor{green!55!black}{$+15.51\%$}} \\
\midrule
SID & FlowRec$^{\dagger}$ & 2.64{\tiny$\pm$0.00} & 4.04{\tiny$\pm$0.03} & 5.91{\tiny$\pm$0.09} & 1.49{\tiny$\pm$0.01} & 1.94{\tiny$\pm$0.01} & 2.41{\tiny$\pm$0.02} \\
 & TIGER & 2.01{\tiny$\pm$0.05} & 3.20{\tiny$\pm$0.03} & 4.62{\tiny$\pm$0.06} & 1.32{\tiny$\pm$0.03} & 1.70{\tiny$\pm$0.02} & 2.06{\tiny$\pm$0.02} \\
 & MaskGR & \underline{3.30}{\tiny$\pm$0.05} & \textbf{5.19}{\tiny$\pm$0.07} & \underline{7.41}{\tiny$\pm$0.07} & \underline{2.14}{\tiny$\pm$0.02} & \underline{2.75}{\tiny$\pm$0.03} & \underline{3.31}{\tiny$\pm$0.03} \\
 & \textbf{\methodname} & \textbf{3.47}{\tiny$\pm$0.02}\,{\scriptsize\textcolor{green!55!black}{$+5.15\%$}} & \underline{5.17}{\tiny$\pm$0.00}\,{\scriptsize\textcolor{red!70!black}{$-0.39\%$}} & \textbf{7.50}{\tiny$\pm$0.04}\,{\scriptsize\textcolor{green!55!black}{$+1.21\%$}} & \textbf{2.29}{\tiny$\pm$0.02}\,{\scriptsize\textcolor{green!55!black}{$+7.01\%$}} & \textbf{2.84}{\tiny$\pm$0.02}\,{\scriptsize\textcolor{green!55!black}{$+3.27\%$}} & \textbf{3.43}{\tiny$\pm$0.02}\,{\scriptsize\textcolor{green!55!black}{$+3.63\%$}} \\
\midrule
PCA & FlowRec & \underline{3.62}{\tiny$\pm$0.02} & \underline{5.54}{\tiny$\pm$0.04} & \underline{8.03}{\tiny$\pm$0.04} & 2.04{\tiny$\pm$0.03} & 2.66{\tiny$\pm$0.04} & 3.28{\tiny$\pm$0.04} \\
 & TIGER & 2.05{\tiny$\pm$0.03} & 3.25{\tiny$\pm$0.03} & 4.61{\tiny$\pm$0.01} & 1.34{\tiny$\pm$0.02} & 1.73{\tiny$\pm$0.02} & 2.08{\tiny$\pm$0.01} \\
 & MaskGR & 3.46{\tiny$\pm$0.03} & 5.25{\tiny$\pm$0.06} & 7.46{\tiny$\pm$0.07} & \underline{2.24}{\tiny$\pm$0.03} & \underline{2.82}{\tiny$\pm$0.02} & \underline{3.38}{\tiny$\pm$0.01} \\
 & \textbf{\methodname} & \textbf{3.79}{\tiny$\pm$0.01}\,{\scriptsize\textcolor{green!55!black}{$+4.70\%$}} & \textbf{5.77}{\tiny$\pm$0.02}\,{\scriptsize\textcolor{green!55!black}{$+4.15\%$}} & \textbf{8.36}{\tiny$\pm$0.08}\,{\scriptsize\textcolor{green!55!black}{$+4.11\%$}} & \textbf{2.34}{\tiny$\pm$0.03}\,{\scriptsize\textcolor{green!55!black}{$+4.46\%$}} & \textbf{2.98}{\tiny$\pm$0.02}\,{\scriptsize\textcolor{green!55!black}{$+5.67\%$}} & \textbf{3.63}{\tiny$\pm$0.04}\,{\scriptsize\textcolor{green!55!black}{$+7.40\%$}} \\
\bottomrule
\end{tabular}
}
\end{table}

\begin{table}[tbp]
\centering
\caption{\textbf{Steam --- full results.} Conventions as in Table~\ref{tab:raw-beauty}.}
\label{tab:raw-steam}
\resizebox{\linewidth}{!}{\providecommand{\pending}[1]{\textcolor{red}{#1}}

\begin{tabular}{llcccccc}
\toprule
\textbf{steam} & \textbf{model} & HR@5 & HR@10 & HR@20 & N@5 & N@10 & N@20 \\
\midrule
scratch & FlowRec & 8.28{\tiny$\pm$0.33} & 13.54{\tiny$\pm$0.31} & 20.61{\tiny$\pm$0.35} & \underline{5.37}{\tiny$\pm$0.30} & \underline{7.06}{\tiny$\pm$0.29} & \underline{8.84}{\tiny$\pm$0.30} \\
 & TIGER & 7.94{\tiny$\pm$0.17} & 13.18{\tiny$\pm$0.24} & \textbf{21.20}{\tiny$\pm$1.59} & 4.90{\tiny$\pm$0.07} & 6.59{\tiny$\pm$0.07} & 8.58{\tiny$\pm$0.38} \\
 & MaskGR & \textbf{8.52}{\tiny$\pm$0.17} & \textbf{14.21}{\tiny$\pm$0.19} & 20.94{\tiny$\pm$0.05} & \textbf{5.58}{\tiny$\pm$0.18} & \textbf{7.40}{\tiny$\pm$0.19} & \textbf{9.11}{\tiny$\pm$0.16} \\
 & \textbf{\methodname} & \underline{8.35}{\tiny$\pm$0.31}\,{\scriptsize\textcolor{red!70!black}{$-2.00\%$}} & \underline{14.15}{\tiny$\pm$0.28}\,{\scriptsize\textcolor{red!70!black}{$-0.42\%$}} & \underline{21.09}{\tiny$\pm$0.20}\,{\scriptsize\textcolor{red!70!black}{$-0.52\%$}} & 5.16{\tiny$\pm$0.17}\,{\scriptsize\textcolor{red!70!black}{$-7.53\%$}} & 7.02{\tiny$\pm$0.16}\,{\scriptsize\textcolor{red!70!black}{$-5.14\%$}} & 8.77{\tiny$\pm$0.14}\,{\scriptsize\textcolor{red!70!black}{$-3.73\%$}} \\
\midrule
SID & FlowRec$^{\dagger}$ & 8.28{\tiny$\pm$0.33} & 13.54{\tiny$\pm$0.31} & 20.61{\tiny$\pm$0.35} & 5.37{\tiny$\pm$0.30} & 7.06{\tiny$\pm$0.29} & 8.84{\tiny$\pm$0.30} \\
 & TIGER & 8.43{\tiny$\pm$0.10} & 13.28{\tiny$\pm$0.63} & 20.46{\tiny$\pm$1.07} & 5.01{\tiny$\pm$0.03} & 6.57{\tiny$\pm$0.19} & 8.36{\tiny$\pm$0.24} \\
 & MaskGR & \underline{8.64}{\tiny$\pm$0.09} & \textbf{14.43}{\tiny$\pm$0.24} & \underline{20.64}{\tiny$\pm$0.23} & \underline{5.54}{\tiny$\pm$0.09} & \textbf{7.40}{\tiny$\pm$0.14} & \underline{8.97}{\tiny$\pm$0.13} \\
 & \textbf{\methodname} & \textbf{8.69}{\tiny$\pm$0.29}\,{\scriptsize\textcolor{green!55!black}{$+0.58\%$}} & \underline{13.94}{\tiny$\pm$0.32}\,{\scriptsize\textcolor{red!70!black}{$-3.40\%$}} & \textbf{20.82}{\tiny$\pm$0.42}\,{\scriptsize\textcolor{green!55!black}{$+0.87\%$}} & \textbf{5.57}{\tiny$\pm$0.19}\,{\scriptsize\textcolor{green!55!black}{$+0.54\%$}} & \underline{7.25}{\tiny$\pm$0.20}\,{\scriptsize\textcolor{red!70!black}{$-2.03\%$}} & \textbf{8.99}{\tiny$\pm$0.23}\,{\scriptsize\textcolor{green!55!black}{$+0.22\%$}} \\
\midrule
PCA & FlowRec & 8.10{\tiny$\pm$0.24} & 13.45{\tiny$\pm$0.11} & 20.70{\tiny$\pm$0.22} & \underline{5.18}{\tiny$\pm$0.23} & 6.88{\tiny$\pm$0.19} & 8.72{\tiny$\pm$0.22} \\
 & TIGER & 8.22{\tiny$\pm$0.31} & 13.58{\tiny$\pm$0.33} & 18.65{\tiny$\pm$0.63} & 5.00{\tiny$\pm$0.07} & 6.72{\tiny$\pm$0.07} & 8.00{\tiny$\pm$0.17} \\
 & MaskGR & \textbf{8.64}{\tiny$\pm$0.00} & \textbf{14.43}{\tiny$\pm$0.05} & \underline{20.97}{\tiny$\pm$0.23} & \textbf{5.57}{\tiny$\pm$0.01} & \textbf{7.42}{\tiny$\pm$0.02} & \textbf{9.08}{\tiny$\pm$0.07} \\
 & \textbf{\methodname} & \underline{8.30}{\tiny$\pm$0.20}\,{\scriptsize\textcolor{red!70!black}{$-3.94\%$}} & \underline{14.28}{\tiny$\pm$0.02}\,{\scriptsize\textcolor{red!70!black}{$-1.04\%$}} & \textbf{21.50}{\tiny$\pm$0.06}\,{\scriptsize\textcolor{green!55!black}{$+2.53\%$}} & 5.07{\tiny$\pm$0.09}\,{\scriptsize\textcolor{red!70!black}{$-8.98\%$}} & \underline{6.99}{\tiny$\pm$0.03}\,{\scriptsize\textcolor{red!70!black}{$-5.80\%$}} & \underline{8.81}{\tiny$\pm$0.04}\,{\scriptsize\textcolor{red!70!black}{$-2.97\%$}} \\
\bottomrule
\end{tabular}
}
\end{table}

\begin{table}[tbp]
\centering
\caption{\textbf{ML-1M --- full results.} Conventions as in Table~\ref{tab:raw-beauty}.}
\label{tab:raw-ml1m}
\resizebox{\linewidth}{!}{\providecommand{\pending}[1]{\textcolor{red}{#1}}

\begin{tabular}{llcccccc}
\toprule
\textbf{ml-1m} & \textbf{model} & HR@5 & HR@10 & HR@20 & N@5 & N@10 & N@20 \\
\midrule
scratch & FlowRec & 15.01{\tiny$\pm$0.12} & 22.73{\tiny$\pm$0.17} & 32.48{\tiny$\pm$0.30} & 9.90{\tiny$\pm$0.17} & 12.38{\tiny$\pm$0.20} & 14.84{\tiny$\pm$0.24} \\
 & TIGER & 12.59{\tiny$\pm$0.40} & 19.03{\tiny$\pm$0.37} & 27.20{\tiny$\pm$0.68} & 8.47{\tiny$\pm$0.28} & 10.54{\tiny$\pm$0.28} & 12.61{\tiny$\pm$0.36} \\
 & MaskGR & \underline{15.64}{\tiny$\pm$0.40} & \underline{23.61}{\tiny$\pm$0.29} & \underline{33.93}{\tiny$\pm$0.17} & \underline{10.23}{\tiny$\pm$0.28} & \underline{12.79}{\tiny$\pm$0.26} & \underline{15.39}{\tiny$\pm$0.21} \\
 & \textbf{\methodname} & \textbf{17.48}{\tiny$\pm$0.27}\,{\scriptsize\textcolor{green!55!black}{$+11.76\%$}} & \textbf{24.80}{\tiny$\pm$0.09}\,{\scriptsize\textcolor{green!55!black}{$+5.04\%$}} & \textbf{34.01}{\tiny$\pm$0.29}\,{\scriptsize\textcolor{green!55!black}{$+0.24\%$}} & \textbf{11.74}{\tiny$\pm$0.19}\,{\scriptsize\textcolor{green!55!black}{$+14.76\%$}} & \textbf{14.10}{\tiny$\pm$0.14}\,{\scriptsize\textcolor{green!55!black}{$+10.24\%$}} & \textbf{16.41}{\tiny$\pm$0.20}\,{\scriptsize\textcolor{green!55!black}{$+6.63\%$}} \\
\midrule
SID & FlowRec$^{\dagger}$ & \underline{15.01}{\tiny$\pm$0.12} & \underline{22.73}{\tiny$\pm$0.17} & \underline{32.48}{\tiny$\pm$0.30} & \underline{9.90}{\tiny$\pm$0.17} & \underline{12.38}{\tiny$\pm$0.20} & \underline{14.84}{\tiny$\pm$0.24} \\
 & TIGER & 12.52{\tiny$\pm$0.70} & 18.91{\tiny$\pm$0.62} & 26.88{\tiny$\pm$0.66} & 8.20{\tiny$\pm$0.31} & 10.25{\tiny$\pm$0.29} & 12.26{\tiny$\pm$0.30} \\
 & MaskGR & 13.04{\tiny$\pm$1.22} & 20.09{\tiny$\pm$1.31} & 29.45{\tiny$\pm$1.75} & 8.43{\tiny$\pm$0.72} & 10.70{\tiny$\pm$0.74} & 13.05{\tiny$\pm$0.85} \\
 & \textbf{\methodname} & \textbf{16.61}{\tiny$\pm$0.31}\,{\scriptsize\textcolor{green!55!black}{$+10.66\%$}} & \textbf{24.25}{\tiny$\pm$0.08}\,{\scriptsize\textcolor{green!55!black}{$+6.69\%$}} & \textbf{33.70}{\tiny$\pm$0.31}\,{\scriptsize\textcolor{green!55!black}{$+3.76\%$}} & \textbf{11.26}{\tiny$\pm$0.15}\,{\scriptsize\textcolor{green!55!black}{$+13.74\%$}} & \textbf{13.73}{\tiny$\pm$0.08}\,{\scriptsize\textcolor{green!55!black}{$+10.90\%$}} & \textbf{16.11}{\tiny$\pm$0.13}\,{\scriptsize\textcolor{green!55!black}{$+8.56\%$}} \\
\midrule
PCA & FlowRec & \underline{15.36}{\tiny$\pm$0.20} & \underline{22.94}{\tiny$\pm$0.04} & \underline{32.68}{\tiny$\pm$0.14} & \underline{10.14}{\tiny$\pm$0.12} & \underline{12.58}{\tiny$\pm$0.05} & \underline{15.03}{\tiny$\pm$0.09} \\
 & TIGER & 12.16{\tiny$\pm$0.31} & 18.43{\tiny$\pm$0.17} & 27.08{\tiny$\pm$0.40} & 8.12{\tiny$\pm$0.18} & 10.12{\tiny$\pm$0.15} & 12.30{\tiny$\pm$0.06} \\
 & MaskGR & 12.58{\tiny$\pm$0.55} & 19.81{\tiny$\pm$0.56} & 29.10{\tiny$\pm$0.39} & 8.27{\tiny$\pm$0.43} & 10.60{\tiny$\pm$0.42} & 12.95{\tiny$\pm$0.39} \\
 & \textbf{\methodname} & \textbf{17.25}{\tiny$\pm$0.13}\,{\scriptsize\textcolor{green!55!black}{$+12.30\%$}} & \textbf{24.71}{\tiny$\pm$0.38}\,{\scriptsize\textcolor{green!55!black}{$+7.72\%$}} & \textbf{33.98}{\tiny$\pm$0.34}\,{\scriptsize\textcolor{green!55!black}{$+3.98\%$}} & \textbf{11.71}{\tiny$\pm$0.11}\,{\scriptsize\textcolor{green!55!black}{$+15.48\%$}} & \textbf{14.11}{\tiny$\pm$0.17}\,{\scriptsize\textcolor{green!55!black}{$+12.16\%$}} & \textbf{16.44}{\tiny$\pm$0.15}\,{\scriptsize\textcolor{green!55!black}{$+9.38\%$}} \\
\bottomrule
\end{tabular}
}
\end{table}

\begin{table}[tbp]
\centering
\caption{\textbf{Yelp-2022 --- full results.} Conventions as in Table~\ref{tab:raw-beauty}.}
\label{tab:raw-yelp}
\resizebox{\linewidth}{!}{\providecommand{\pending}[1]{\textcolor{red}{#1}}

\begin{tabular}{llcccccc}
\toprule
\textbf{yelp} & \textbf{model} & HR@5 & HR@10 & HR@20 & N@5 & N@10 & N@20 \\
\midrule
scratch & FlowRec & 2.79{\tiny$\pm$0.02} & 4.76{\tiny$\pm$0.05} & 7.76{\tiny$\pm$0.07} & 1.73{\tiny$\pm$0.02} & 2.37{\tiny$\pm$0.02} & 3.12{\tiny$\pm$0.03} \\
 & TIGER & 0.26{\tiny$\pm$0.01} & 0.42{\tiny$\pm$0.02} & 0.73{\tiny$\pm$0.10} & 0.17{\tiny$\pm$0.01} & 0.22{\tiny$\pm$0.01} & 0.30{\tiny$\pm$0.03} \\
 & MaskGR & \underline{2.99}{\tiny$\pm$0.04} & \underline{5.10}{\tiny$\pm$0.07} & \underline{8.24}{\tiny$\pm$0.10} & \underline{1.87}{\tiny$\pm$0.03} & \underline{2.54}{\tiny$\pm$0.03} & \underline{3.33}{\tiny$\pm$0.04} \\
 & \textbf{\methodname} & \textbf{3.07}{\tiny$\pm$0.02}\,{\scriptsize\textcolor{green!55!black}{$+2.68\%$}} & \textbf{5.14}{\tiny$\pm$0.02}\,{\scriptsize\textcolor{green!55!black}{$+0.78\%$}} & \textbf{8.26}{\tiny$\pm$0.02}\,{\scriptsize\textcolor{green!55!black}{$+0.24\%$}} & \textbf{1.93}{\tiny$\pm$0.02}\,{\scriptsize\textcolor{green!55!black}{$+3.21\%$}} & \textbf{2.60}{\tiny$\pm$0.01}\,{\scriptsize\textcolor{green!55!black}{$+2.36\%$}} & \textbf{3.38}{\tiny$\pm$0.01}\,{\scriptsize\textcolor{green!55!black}{$+1.50\%$}} \\
\midrule
SID & FlowRec$^{\dagger}$ & \underline{2.79}{\tiny$\pm$0.02} & \underline{4.76}{\tiny$\pm$0.05} & \underline{7.76}{\tiny$\pm$0.07} & \underline{1.73}{\tiny$\pm$0.02} & \underline{2.37}{\tiny$\pm$0.02} & \underline{3.12}{\tiny$\pm$0.03} \\
 & TIGER & 1.27{\tiny$\pm$0.08} & 2.05{\tiny$\pm$0.10} & 2.95{\tiny$\pm$0.18} & 0.77{\tiny$\pm$0.05} & 1.02{\tiny$\pm$0.06} & 1.25{\tiny$\pm$0.08} \\
 & MaskGR & 2.63{\tiny$\pm$0.01} & 4.37{\tiny$\pm$0.05} & 6.79{\tiny$\pm$0.07} & 1.65{\tiny$\pm$0.01} & 2.20{\tiny$\pm$0.02} & 2.81{\tiny$\pm$0.03} \\
 & \textbf{\methodname} & \textbf{3.13}{\tiny$\pm$0.02}\,{\scriptsize\textcolor{green!55!black}{$+12.19\%$}} & \textbf{5.18}{\tiny$\pm$0.02}\,{\scriptsize\textcolor{green!55!black}{$+8.82\%$}} & \textbf{8.32}{\tiny$\pm$0.01}\,{\scriptsize\textcolor{green!55!black}{$+7.22\%$}} & \textbf{1.99}{\tiny$\pm$0.01}\,{\scriptsize\textcolor{green!55!black}{$+15.03\%$}} & \textbf{2.64}{\tiny$\pm$0.01}\,{\scriptsize\textcolor{green!55!black}{$+11.39\%$}} & \textbf{3.43}{\tiny$\pm$0.01}\,{\scriptsize\textcolor{green!55!black}{$+9.94\%$}} \\
\midrule
PCA & FlowRec & \underline{2.98}{\tiny$\pm$0.05} & \underline{5.10}{\tiny$\pm$0.05} & \underline{8.34}{\tiny$\pm$0.06} & \underline{1.86}{\tiny$\pm$0.03} & \underline{2.54}{\tiny$\pm$0.03} & \underline{3.35}{\tiny$\pm$0.04} \\
 & TIGER & 1.29{\tiny$\pm$0.07} & 2.09{\tiny$\pm$0.09} & 3.03{\tiny$\pm$0.10} & 0.80{\tiny$\pm$0.04} & 1.06{\tiny$\pm$0.04} & 1.30{\tiny$\pm$0.04} \\
 & MaskGR & 2.53{\tiny$\pm$0.02} & 4.26{\tiny$\pm$0.04} & 6.60{\tiny$\pm$0.04} & 1.59{\tiny$\pm$0.01} & 2.14{\tiny$\pm$0.01} & 2.73{\tiny$\pm$0.02} \\
 & \textbf{\methodname} & \textbf{3.18}{\tiny$\pm$0.02}\,{\scriptsize\textcolor{green!55!black}{$+6.71\%$}} & \textbf{5.25}{\tiny$\pm$0.04}\,{\scriptsize\textcolor{green!55!black}{$+2.94\%$}} & \textbf{8.38}{\tiny$\pm$0.05}\,{\scriptsize\textcolor{green!55!black}{$+0.48\%$}} & \textbf{2.03}{\tiny$\pm$0.02}\,{\scriptsize\textcolor{green!55!black}{$+9.14\%$}} & \textbf{2.70}{\tiny$\pm$0.03}\,{\scriptsize\textcolor{green!55!black}{$+6.30\%$}} & \textbf{3.48}{\tiny$\pm$0.03}\,{\scriptsize\textcolor{green!55!black}{$+3.88\%$}} \\
\bottomrule
\end{tabular}
}
\end{table}

\begin{table}[tbp]
\centering
\caption{\textbf{Amazon Office --- full results.} Conventions as in Table~\ref{tab:raw-beauty}.}
\label{tab:raw-office}
\resizebox{\linewidth}{!}{\providecommand{\pending}[1]{\textcolor{red}{#1}}

\begin{tabular}{llcccccc}
\toprule
\textbf{office} & \textbf{model} & HR@5 & HR@10 & HR@20 & N@5 & N@10 & N@20 \\
\midrule
scratch & FlowRec & 2.11{\tiny$\pm$0.06} & 3.09{\tiny$\pm$0.07} & 4.33{\tiny$\pm$0.12} & 1.40{\tiny$\pm$0.03} & 1.71{\tiny$\pm$0.03} & 2.03{\tiny$\pm$0.05} \\
 & TIGER & 0.76{\tiny$\pm$0.01} & 1.09{\tiny$\pm$0.02} & 1.67{\tiny$\pm$0.10} & 0.48{\tiny$\pm$0.02} & 0.59{\tiny$\pm$0.02} & 0.74{\tiny$\pm$0.02} \\
 & MaskGR & \underline{2.29}{\tiny$\pm$0.02} & \underline{3.40}{\tiny$\pm$0.03} & \textbf{4.88}{\tiny$\pm$0.03} & \underline{1.53}{\tiny$\pm$0.01} & \underline{1.89}{\tiny$\pm$0.02} & \underline{2.26}{\tiny$\pm$0.01} \\
 & \textbf{\methodname} & \textbf{2.72}{\tiny$\pm$0.01}\,{\scriptsize\textcolor{green!55!black}{$+18.78\%$}} & \textbf{3.50}{\tiny$\pm$0.01}\,{\scriptsize\textcolor{green!55!black}{$+2.94\%$}} & \underline{4.49}{\tiny$\pm$0.01}\,{\scriptsize\textcolor{red!70!black}{$-7.99\%$}} & \textbf{2.00}{\tiny$\pm$0.01}\,{\scriptsize\textcolor{green!55!black}{$+30.72\%$}} & \textbf{2.25}{\tiny$\pm$0.01}\,{\scriptsize\textcolor{green!55!black}{$+19.05\%$}} & \textbf{2.50}{\tiny$\pm$0.01}\,{\scriptsize\textcolor{green!55!black}{$+10.62\%$}} \\
\midrule
SID & FlowRec$^{\dagger}$ & 2.11{\tiny$\pm$0.06} & 3.09{\tiny$\pm$0.07} & 4.33{\tiny$\pm$0.12} & 1.40{\tiny$\pm$0.03} & 1.71{\tiny$\pm$0.03} & 2.03{\tiny$\pm$0.05} \\
 & TIGER & 1.47{\tiny$\pm$0.07} & 2.26{\tiny$\pm$0.11} & 3.17{\tiny$\pm$0.16} & 0.95{\tiny$\pm$0.04} & 1.20{\tiny$\pm$0.05} & 1.43{\tiny$\pm$0.07} \\
 & MaskGR & \underline{2.39}{\tiny$\pm$0.01} & \underline{3.56}{\tiny$\pm$0.04} & \underline{4.88}{\tiny$\pm$0.05} & \underline{1.59}{\tiny$\pm$0.02} & \underline{1.96}{\tiny$\pm$0.02} & \underline{2.30}{\tiny$\pm$0.03} \\
 & \textbf{\methodname} & \textbf{2.95}{\tiny$\pm$0.03}\,{\scriptsize\textcolor{green!55!black}{$+23.43\%$}} & \textbf{4.13}{\tiny$\pm$0.05}\,{\scriptsize\textcolor{green!55!black}{$+16.01\%$}} & \textbf{5.65}{\tiny$\pm$0.07}\,{\scriptsize\textcolor{green!55!black}{$+15.78\%$}} & \textbf{2.07}{\tiny$\pm$0.02}\,{\scriptsize\textcolor{green!55!black}{$+30.19\%$}} & \textbf{2.45}{\tiny$\pm$0.02}\,{\scriptsize\textcolor{green!55!black}{$+25.00\%$}} & \textbf{2.84}{\tiny$\pm$0.03}\,{\scriptsize\textcolor{green!55!black}{$+23.48\%$}} \\
\midrule
PCA & FlowRec & \underline{2.90}{\tiny$\pm$0.02} & \textbf{4.41}{\tiny$\pm$0.04} & \textbf{6.23}{\tiny$\pm$0.02} & \underline{1.67}{\tiny$\pm$0.02} & \underline{2.15}{\tiny$\pm$0.02} & \underline{2.61}{\tiny$\pm$0.02} \\
 & TIGER & 1.52{\tiny$\pm$0.01} & 2.33{\tiny$\pm$0.03} & 3.34{\tiny$\pm$0.08} & 0.98{\tiny$\pm$0.01} & 1.25{\tiny$\pm$0.00} & 1.50{\tiny$\pm$0.02} \\
 & MaskGR & 2.16{\tiny$\pm$0.02} & \underline{3.24}{\tiny$\pm$0.01} & 4.48{\tiny$\pm$0.01} & 1.42{\tiny$\pm$0.01} & 1.77{\tiny$\pm$0.01} & 2.08{\tiny$\pm$0.01} \\
 & \textbf{\methodname} & \textbf{3.22}{\tiny$\pm$0.01}\,{\scriptsize\textcolor{green!55!black}{$+11.03\%$}} & \textbf{4.41}{\tiny$\pm$0.01}\,{\scriptsize$0.00\%$} & \underline{5.84}{\tiny$\pm$0.03}\,{\scriptsize\textcolor{red!70!black}{$-6.26\%$}} & \textbf{2.28}{\tiny$\pm$0.01}\,{\scriptsize\textcolor{green!55!black}{$+36.53\%$}} & \textbf{2.66}{\tiny$\pm$0.01}\,{\scriptsize\textcolor{green!55!black}{$+23.72\%$}} & \textbf{3.02}{\tiny$\pm$0.01}\,{\scriptsize\textcolor{green!55!black}{$+15.71\%$}} \\
\bottomrule
\end{tabular}
}
\end{table}

\end{document}